\documentclass[aps,showpacs,floatfix,superscriptaddress,12pt,oneside,headinclude,footinclude,prb]{revtex4-1}
\usepackage{amsmath}
\usepackage{latexsym}
\usepackage{geometry}
\usepackage[english]{babel}
\usepackage{blindtext}
\usepackage{graphicx}
\usepackage{subfigure}
\usepackage{indentfirst}
\usepackage{appendix}
\usepackage{mathrsfs}
\usepackage{color}
\usepackage[colorlinks,linkcolor=blue,anchorcolor=blue,citecolor=green]{hyperref}

\graphicspath{{eps/}}

\begin{document}

\title{Superconductivity at the three-dimensional Anderson metal-insulator transition}
\author{Bo Fan}
\email{bo.fan@sjtu.edu.cn}
\affiliation{Shanghai Center for Complex Physics, 
	School of Physics and Astronomy, Shanghai Jiao Tong
	University, Shanghai 200240, China}
\author{Antonio M. Garc\'ia-Garc\'ia}
\email{amgg@sjtu.edu.cn}
\affiliation{Shanghai Center for Complex Physics, 
	School of Physics and Astronomy, Shanghai Jiao Tong
	University, Shanghai 200240, China}

\begin{abstract}We study a disordered weakly-coupled superconductor around the Anderson transition by solving numerically the Bogoliubov-de Gennes (BdG) equations in a three dimensional lattice of size up to $20\times20\times20$ in the presence of a random potential. The spatial average of the order parameter is moderately enhanced as disorder approaches the transition but decreases sharply in the insulating region. The spatial distribution of the order parameter is sensitive to the disorder strength: for intermediate disorders below the transition, we already observe a highly asymmetric distribution with an exponential tail. Around the transition, it is well described by a log-normal distribution and a parabolic singularity spectrum. These features are typical of a multifractal measure. We determine quantitatively the critical disorder at which the insulator transition occurs by an analysis of level statistics in the spectral region that contributes to the formation of the order parameter. Interestingly, spectral correlations at the transition are similar to those found in non-interacting disordered systems at the Anderson transition. A percolation analysis suggests that the loss of phase coherence may occur around the critical disorder.   
\end{abstract}\maketitle

\newpage
\section{Introduction}
Quantum coherence effects are of paramount importance in the dynamics of disordered and quantum chaotic systems. 
However, until rather recently, its effect on superconductivity has been relatively overlooked. A reason for that is the so called Anderson theorem \cite{Anderson1959}, also postulated by Gorkov \cite{Gorkov1961}, that non-magnetic impurities in metals did not break Cooper pairs and therefore have only a relatively small effect on superconductivity. In parallel, experiments in metallic superconductors \cite{Abeles1966,Abeles1967} were relatively well described without the need to consider these effects. 
However, computational advances together with an enhanced experimental control and the introduction of the scanning tunneling microscope started to reveal a completely different picture. Numerical solutions of two dimensional BdG equations in a random potential \cite{ghosal1998,Ghosal2001} showed an emergent granularity and strong spatial fluctuations of the order parameter even for disorder strengths within the metallic region but not far from superconductor-insulator transition. This emergent granularity was later corroborated experimentally \cite{goldman1993,Orr1985,Jaeger1986,Jaeger1989,Sacepe2011,Trivedi2012,Mondal2011,Lemarie2013,Chand2012}. Indeed, as spatial dimensionality is reduced, it was explicitly observed that quantum coherence effects became increasingly relevant \cite{xue2010,Uchihashi2016}. For instance, quantum size effects related to confinement were predicted theoretically \cite{Blatt1963,Parmenter1968a,Shanenko2006,Shanenko2007,Garcia-Garcia2008,Garcia-Garcia2011} and later confirmed experimentally in Sn and Pb superconducting nano-grains \cite{Bose2010,Brihuega2011}. 

A distinct feature of the interplay of quantum coherence and disorder in the non-interacting limit is the multifractality of eigenstates \cite{Castellani1986,Wegner1980,Falko1995} that occurs around the mobility edge separating metallic and insulating states in three and higher dimensions \cite{abraham1979}. 
Two dimensions (2D) is the critical dimension \cite{abraham1979} for localization. Strictly speaking, in an infinity disordered two dimensional system, all states are exponentially localized. However, for weak disorder, the localization length is exponentially large and, for smaller sizes, the system shows multifractal features \cite{Wegner1980,Falko1995} in a relatively large window of disorder strengths. Moreover, other effects such as spin orbit-interaction may induce a transition strictly in two dimensions \cite{ando1989}. 

The interplay between weak multifractality and superconductivity in two dimensions was recently studied \cite{Mayoh2015} using a simple Bardeen-Cooper-Schrieffer (BCS) formalism that assumed that the order parameter was well described by the multifractal eigenstates of the one-body problem. It was found that the spatial distribution of the order parameter is described by a log-normal distribution. The spatial average of the distribution increases with  disorder and it can be substantially larger than the order parameter in the clean limit. The qualitative effect of Coulomb interactions in this critical region, investigated earlier \cite{mirlin2013}, predicted a much dramatic enhancement.   
Recent experiments \cite{verdu2018,xue2019} in weakly disordered two dimensional NbSe$_2$ and theoretical results based on the numerical solution of the BdG equations \cite{Bofan2020,Gastiasoro2018} have confirmed both the enhancement of superconductivity with disorder and the log-normal distribution of the order parameter.

In three dimensions (3D), the Anderson transition occurs for strong disorder which makes more difficult a theoretical treatment due to the absence of a small parameter. The interplay between the Anderson transition and superconductivity, was first investigated in Refs.\cite{Feigelman2007,Feigelman2010}, earlier than the two dimensional analysis mentioned above, by using also a BCS approach. According to their analysis, the order parameter is enhanced dramatically, up to orders of magnitude with respect to the clean limit, and its moments \cite{Feigelman2010} are consistent with those of a log-normal distribution. So far, experiments could not reproduce these features. 

Here we compare these expectations with the outcome of the full numerical solution of the 3D BdG equations for different disorder strengths with an especial emphasis in the region around the superconductor-insulator transition. While the spatial average of the order parameter increases moderately with disorder, this increasing stops when the system approaches the transition. The spatial distribution of the order parameter becomes increasingly broad even for disorder strength far from the transition. Around the transition, it is close to log-normal as in the 2D case \cite{mayoh2015global}.

The critical disorder is determined by the analysis of level statistics \cite{shapiro1993,altshuler1988} in the spectral region that contributes to the buildup of the order parameter. Spectral correlations around the transition are intermediate between those of a metal and insulator and qualitatively similar to those \cite{shapiro1993} of a non-interacting disordered metal at the Anderson transition. The disorder strength at which phase coherence is lost, estimated by a percolation analysis, is similar to that at which the superconductor-insulator transition occurs.

The paper is organized as follows. 
In section \ref{sec:model}, we introduce the model and determine the range of parameters where our calculation is reliable. In section \ref{sec:enhancesc}, we compute numerically the spatial average of the order parameter $\langle \Delta(r) \rangle$, and determine the range of parameters for which enhancement of superconductivity occurs. The dependence of disorder of the local density of states is the subject of section \ref{sec:dosdis}.
Section \ref{sec:gapdistribution} is devoted to the study of the 
 spatial distribution, and the singularity spectrum of the order parameter. 
 In section \ref{sec:overlap}, we compute the overlap of eigenstates which allows us to estimate the effective spectral window around the Fermi energy which contributes significantly to the formation of the order parameter. In section \ref{sec:levelstatistics}, we estimate the critical disorder at which the superconductor-insulator transition occurs by an analysis of level statistics. We also show that level statistics around the transition is intermediate between Poisson statistics and random matrix theory as in a non-interacting disordered system at the Anderson transition. In section \ref{sec:percolation}, we carry out a percolation analysis in order to estimate the disorder strength at which phase coherence is lost. We find that the percolating transition occurs around the same disorder as the metal-insulator transition. In section \ref{sec:discussion}, we summarize the main findings of the paper and enumerate a few related problems for future research.

\section{Disordered Bogoliubov-de Gennes equations}\label{sec:model}
The following BdG equations \cite{Ghosal2001,DeGennes1964,DeGennes1966} result from the evaluation of the path integral of a disordered fermionic tight binding model in a cubic lattice with short-range attractive
interactions by the saddle-point method that is only exact in the mean-field limit: 

\begin{equation}
\left(\begin{matrix}
\hat{K} 		& \hat{\Delta}  \\
\hat{\Delta}^* 	& -\hat{K}^* 	\\
\end{matrix}\right)
\left(\begin{matrix}
u_n(r_i)   \\
v_n(r_i) \\
\end{matrix}\right)
= E_n
\left(\begin{matrix}
u_n(r_i)   \\
v_n(r_i) \\
\end{matrix}\right)
\label{eq.1}
\end{equation} 

where
\begin{equation}
\hat{K}u_n(r_i)=-t\sum_{\delta}u_n(r_i+\delta)+(V_i-\mu_i)u_n(r_i),
\label{eq.2}
\end{equation}

$\delta$ stands for the nearest neighboring sites, $t$ is the hopping strength, $V_i$ is strength of the random potential at site $i$,  extracted from an uniform distribution  $[-V/2,V/2]$, $\mu_i = \mu + |U|n(r_i)/2$ incorporates the site-dependent Hartree shift. The chemical potential $\mu$, is determined by the averaged density $\langle n \rangle =\sum_{i}n(r_i)/N$. $U$ is the pairing interaction, and $\hat{\Delta}u_n(r_i) \equiv \Delta(r_i)u_n(r_i)$. The same definition applies to $v_n(r_i)$. The BdG equations are completed by the self-consistency conditions for the site dependent order parameter $\Delta(r_i)$ and density $n(r_i)$,
\begin{equation}
\Delta(r_i) = |U|\sum_{E_n\leq \omega_D}u_n(r_i)v_n^*(r_i)
\label{eq.3}
\end{equation}
and 
\begin{equation}
n(r_i) =2\sum_{n}|v_n(r_i)|^2,
\label{eq.4}
\end{equation}
where $\omega_D$ is the cut-off energy. 
We solve these equations for a cubic lattice of $N = L\times L\times L$ sites, where $L$ is the side length of the sample in units of the lattice constant. In order to minimize finite size effects, we employ the periodic boundary conditions. We employ a standard iterative algorithm. Starting with an initial seed for the order parameter, we solve Eq.~\eqref{eq.1} numerically, and obtain the eigenvalues ${E_n}$ and the corresponding eigenvectors $\{u_n(r_i),v_n(r_i)\}$. We then use the self-consistent condition, Eqs.~\eqref{eq.3} and \eqref{eq.4}, to get the new value of ${\Delta(r_i)}$ and ${\mu_i}$. We repeat the process until the absolute error of ${\Delta(r_i)}$ is smaller than $5\times10^{-6}$ or the relative error is smaller than $1\times10^{-3}$. For convenience, all the parameters are in units of $t=1$ and the density is fixed at  $\langle n \rangle = 0.875$ throughout the paper. 

\subsection{Characteristic superconducting length and choice of parameters}
\begin{figure}
	\begin{center}
		\includegraphics[width=10cm]{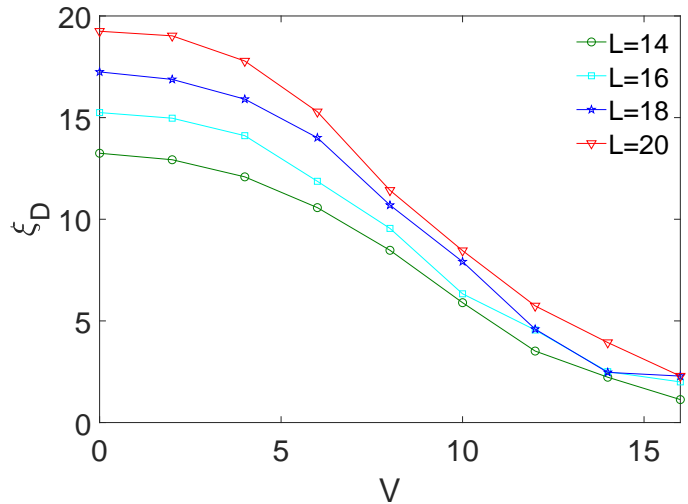}
		\caption{The characteristic length $\xi_D$ as a function of disorder for $U = -1$ that indicates the typical size of a superconductor island.  $\xi_D$ changes slowly when disorder is weak $V\leq6$. However, for $8 \leq V\leq12$, $\xi_D$ decreases faster and is much smaller than the system size which assures the reliability of our numerical results.}\label{Fig.xi}
	\end{center}
\end{figure}

Our first task is to determine the range of parameters where our calculation is reliable. For this to happen, the typical length of the superconducting state must be smaller than the system size. For the former, we choose the typical size of the order of the parameter correlations $\xi_D$, 
\begin{equation}
\xi_D = \sqrt{\frac{\sum_{r}\langle\Delta(0)\Delta(r)\rangle r^2}{N\langle\Delta(0)\Delta(0)\rangle}}
\label{eq.5}
\end{equation}
which is close to the standard superconducting coherence length. 
As we mentioned earlier, the quantum coherence effects we aim to investigate are stronger if the electron-phonon coupling $U$ is weaker. Therefore, we set $|U|$ to the smallest possible value so that $\xi_D$ is less than the maximum size $L \sim 20$ we can reach numerically in the region of relatively strong disorder, close to the transition, we are mostly interested in. 
The results shown in Fig.~\ref{Fig.xi} indicate that $U = -1$ is the smallest coupling for which we can obtain reliable results. In the weak disorder region $V \leq 6$, $\xi_D$ is almost the system size but for stronger disorder $V \sim 10$, $\xi_D$ is reduced considerable so finite size effects are not important and our results are reliable in this region. We note that the dimensionless coupling constant $\lambda$ increases with $|U|$, and also with $\langle n \rangle$, and our choice of couplings is close to that of realistic weakly coupled metallic superconductors such as Sn.

\section{Spatial average of $\langle \Delta(r_i) \rangle$ and enhancement of superconductivity by disorder}\label{sec:enhancesc}
\begin{figure}
	\begin{center}
		\subfigure[]{\label{fig.gap_1}
			\includegraphics[width=8.5cm]{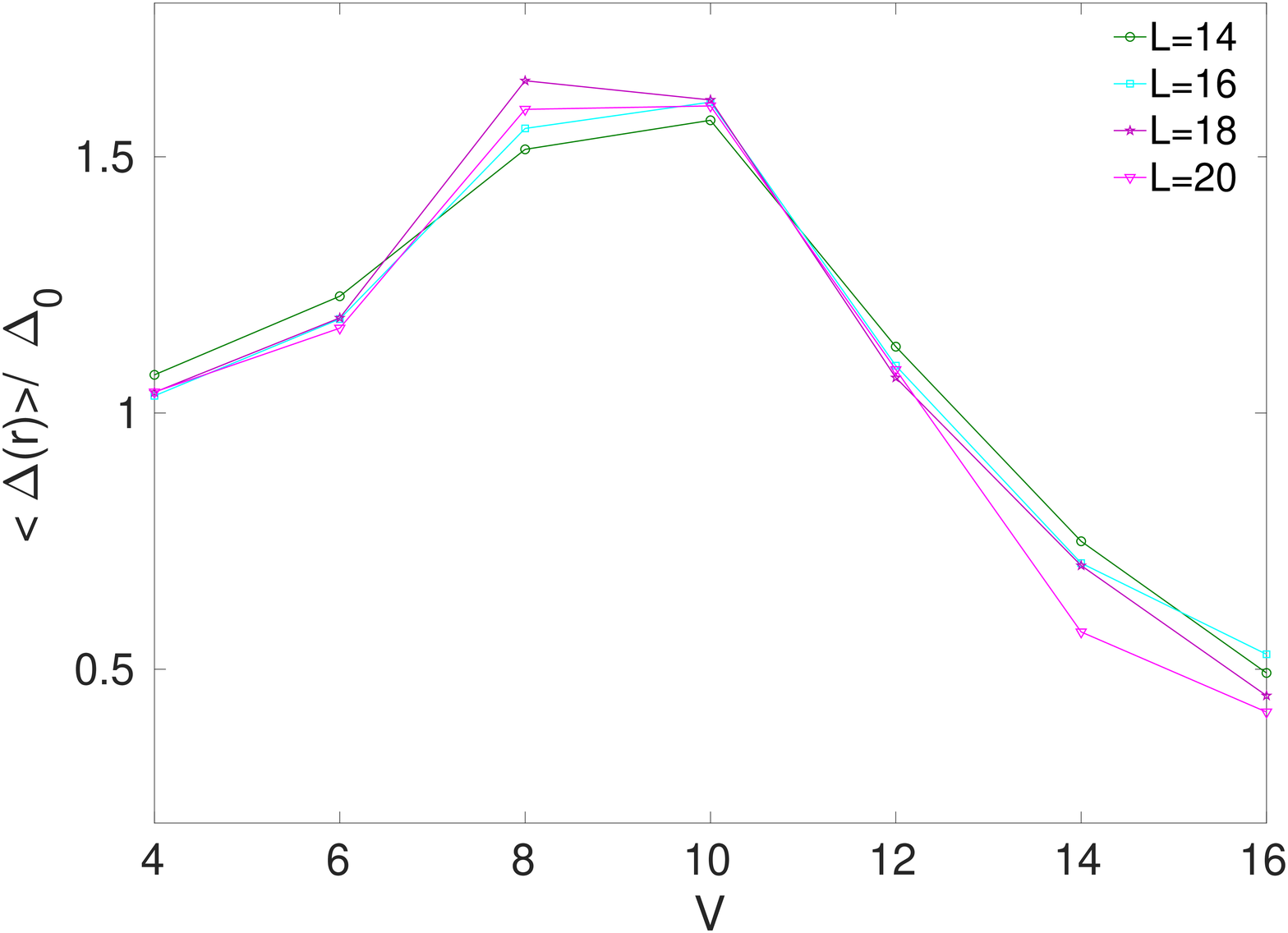}}
		\subfigure[]{\label{fig.gap_2}
			\includegraphics[width=8.5cm]{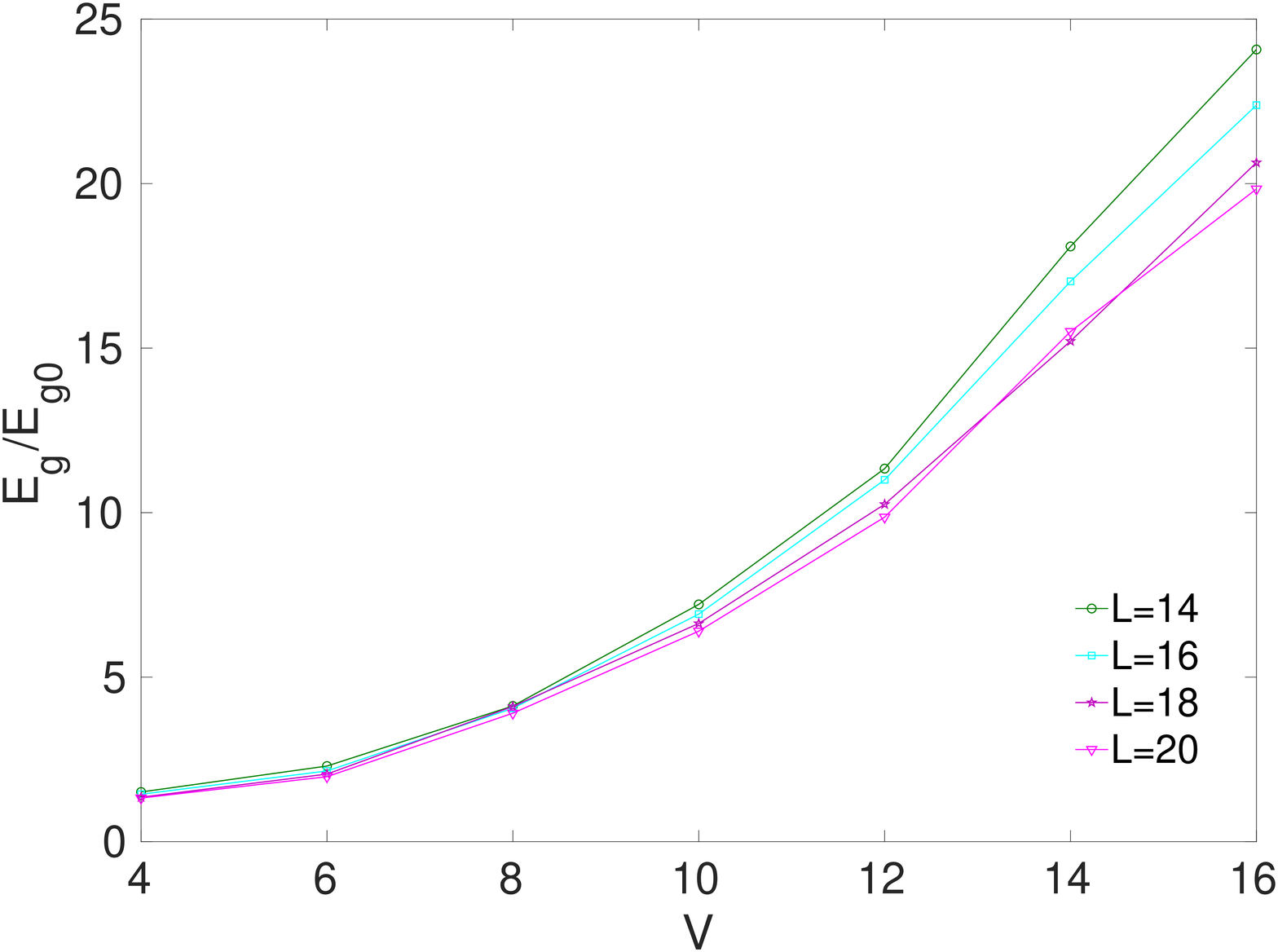}}
		\caption{The spatial average of the order parameter $\langle \Delta(r) \rangle$ (Normalized by $\Delta_0 \sim 0.002t$) and the spectral gap $E_{g}$ (Normalized by $E_{g0} \sim 0.002t$), obtained from the solution of the BdG equations, as a function of disorder $V$ for different sizes and $U = -1$. For weak disorder $V \textless 4$, size effects are rather large (not shown) indicating that the sample size is not large enough to get meaningful results. Therefore we restrict our analysis to $V\geq4$ where size effects are not important.
		The numerical results $\langle \Delta(r) \rangle$ are in agreement with the analytical prediction of Ref. \cite{Mayoh2015}, based on a simpler BCS approach, the average order parameter increases with disorder which suggests that disorder can enhance superconductivity. Finally, it decreases in the strong disorder regime. We shall see that the latter is due to the weakening of eigenstates overlap close to the Fermi energy. By contrast, as in the 2D case, the spectral gap increases with disorder monotonically. 
		}\label{Fig.gap}
	\end{center}
\end{figure}

We compute the disorder dependence of the spatial average of the order parameter $\langle \Delta(r) \rangle = 1/N\sum_{i}\Delta(r_i)$ in order to clarify whether the amplitude of the order parameter is enhanced by disorder. We have found that, see Fig.~\ref{Fig.gap}, the averaged order parameter $\langle \Delta(r) \rangle$ indeed increases with disorder though this increase eventually stops for $V \sim 12$. For stronger disorder, it decreases monotonically. We shall see that the maximum occurs around the critical region where the transition occurs.  For very weak disorder $V \sim 2$ (not shown), where our calculation is less reliable, we observe a decrease of the order parameter with respect to the clean limit which is likely a finite size effect of no much relevance in this context as it will be severely reduced if the system size could be increased. 
These results are different from the analytical \cite{Mayoh2014a} and numerical results \cite{Bofan2020,Gastiasoro2018} in the two dimensional weak-coupling, weak-disorder limit where the enhancement is substantially larger and no decrease for stronger disorder was observed. Although these features may depend on the coupling strength, the differences are ultimately related to the fact that, in two dimensions, the effective critical region is much broader. These results seem also in disagreement with previous BCS analytical results \cite{Feigelman2007,Feigelman2010} at the three dimensional transition where the predicted enhancement of the order parameter with disorder is much larger as the order parameter has a power-law dependence with the dimensionless electron-phonon coupling. 

For the sake of completeness, we also compute the energy gap $E_{g}$. We observe, see Fig.~\ref{fig.gap_2}, a monotonic increase with disorder that agrees with the average of the order parameter in the weak disordered limit only. This discrepancy between the two quantities for sufficiently strong disorder is also observed in 2D disordered superconductors \cite{ghosal1998,Bofan2020,Ghosal2001}.

As in the two dimensional case, the increase for strong disorder in the insulating region is a consequence of Anderson localization effects that enlarge the mean level spacing as the typical distance is no longer the system size, but the localization length that decreases as disorder increases. Therefore, the observed monotonous increase with disorder, that does not flatten or reverse tendency around the transition, is not related to superconductivity for sufficiently strong disorder but rather with the physics of Anderson localization. In summary, disorder in three dimensions may enhance superconductivity but it is a relatively small effect that stops around the critical region. On the insulating side, disorder is always detrimental of superconductivity. 

\section{Density of states}\label{sec:dosdis}

\begin{figure}
	\begin{center}
		\includegraphics[width=10cm]{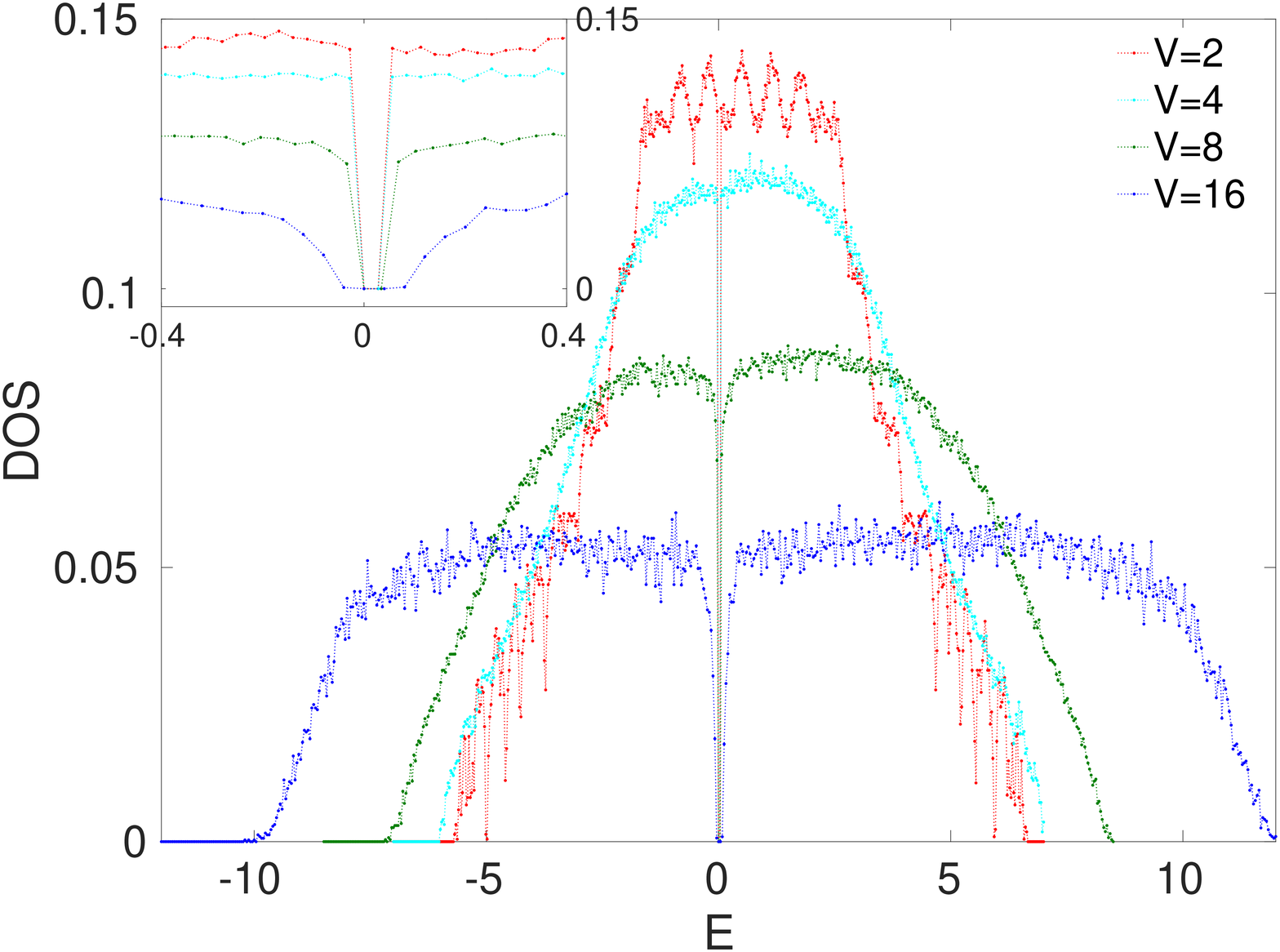}
		\caption{Density of states (DOS) for different disorder ($V=2, 4, 8$ and $16$). The inset is the DOS between $-0.4\leq E \leq0.4$ to show the gap. Disorder reduces the DOS, but enlarge the energy band and the gap around $E=0$. As in the non-interacting case, the DOS varies smoothly with disorder and therefore it is not a good indicator of the Anderson transition.}\label{Fig.dos}
	\end{center}
\end{figure}
In this section we investigate the impact of disorder in the local density of states (DOS), 
\begin{equation}
DOS=\frac{1}{N}\sum_{r_i} [u_n^2(r_i)\delta(E-E_n)+v_n^2(r_i)\delta(E+E_n)]
\end{equation} 
aimed to illustrate similarities and differences with the non-interacting case.  
There is always a finite gap around $E=0$, see the inset in Fig.~\ref{Fig.dos}, representing the superconducting energy gap. The DOS have two peaks around the gap corresponding to the superconducting coherence peaks, a signature of BCS theory. These peaks are suppressed in the 2D strong disorder limit \cite{Ghosal2001}. Other features are qualitatively similar to that of the non-interacting limit \cite{Markos2006}. For instance, for weak disorder, we observe that, as in the non-interacting limit,  oscillations eventually vanish as disorder increases. Likewise, the DOS is reduced for stronger disorder but the spectral support increases. These similarities suggest that, at least in the weak disorder regime, where coherence effects are not important, the eigenstates of the BdG equations may be qualitatively similar to those in the non-interacting limit which may justify a BCS approach at least for not too strong disorder.
 
Finally, we note the spectrum of the BdG equations has a {\it parity} symmetry in the non-interacting limit $|U| \to 0$ \cite{Markos2006}, namely, $DOS(E)=DOS(-E)$. However, once interactions are switched on, the spectrum of the BdG equations, and therefore the related DOS, does not have this symmetry. As a consequence, the spectrum is effectively shifted. We know that in the non-interacting case, the wave function corresponding to $E = 0$ is always the most extended state in comparison with other energies. If the spectrum is shifted, the wave function $u(r)$ and $v(r)$ around $E=0$ are no longer the most extended states. However, only states around $E=0$ contribute to the order parameter significantly. Therefore, this shift in the DOS may explain why the critical disorder is smaller in the BdG equations with respect to the non-interacting limit.

\section{Spatial distribution of the order parameter} \label{sec:gapdistribution}

In this section, we investigate the spatial dependence of the amplitude of the order parameter $\Delta(r_i)$. Our main motivation is to characterize its spatial distribution as a function of disorder. Of special interest is to clarify the role of the log-normal spatial distribution \cite{Mayoh2015,Bofan2020} that describes the distribution of $\Delta(r_i)$ of two dimensional, weakly-coupled, weakly-disordered superconductors. The analytical derivation of the log-normal distribution \cite{Mayoh2015} in the 2D case is heavily based on the assumption of weak disorder, large conductance, so it is unclear to be valid at the Anderson transition in three dimensions where disorder is strong and the dimensionless conductance is of order one. We also analyze the singularity spectrum $f(\alpha)$ \cite{Jensen1989} to obtain further information of the spatial distribution of the order parameter around the transition.

\subsection{Spatial dependence and probability distribution of the order parameter amplitude} 
\begin{figure}[htbp]
	\begin{center}
		\subfigure[]{\label{fig.spatial_1} 
			\includegraphics[width=8.5cm]{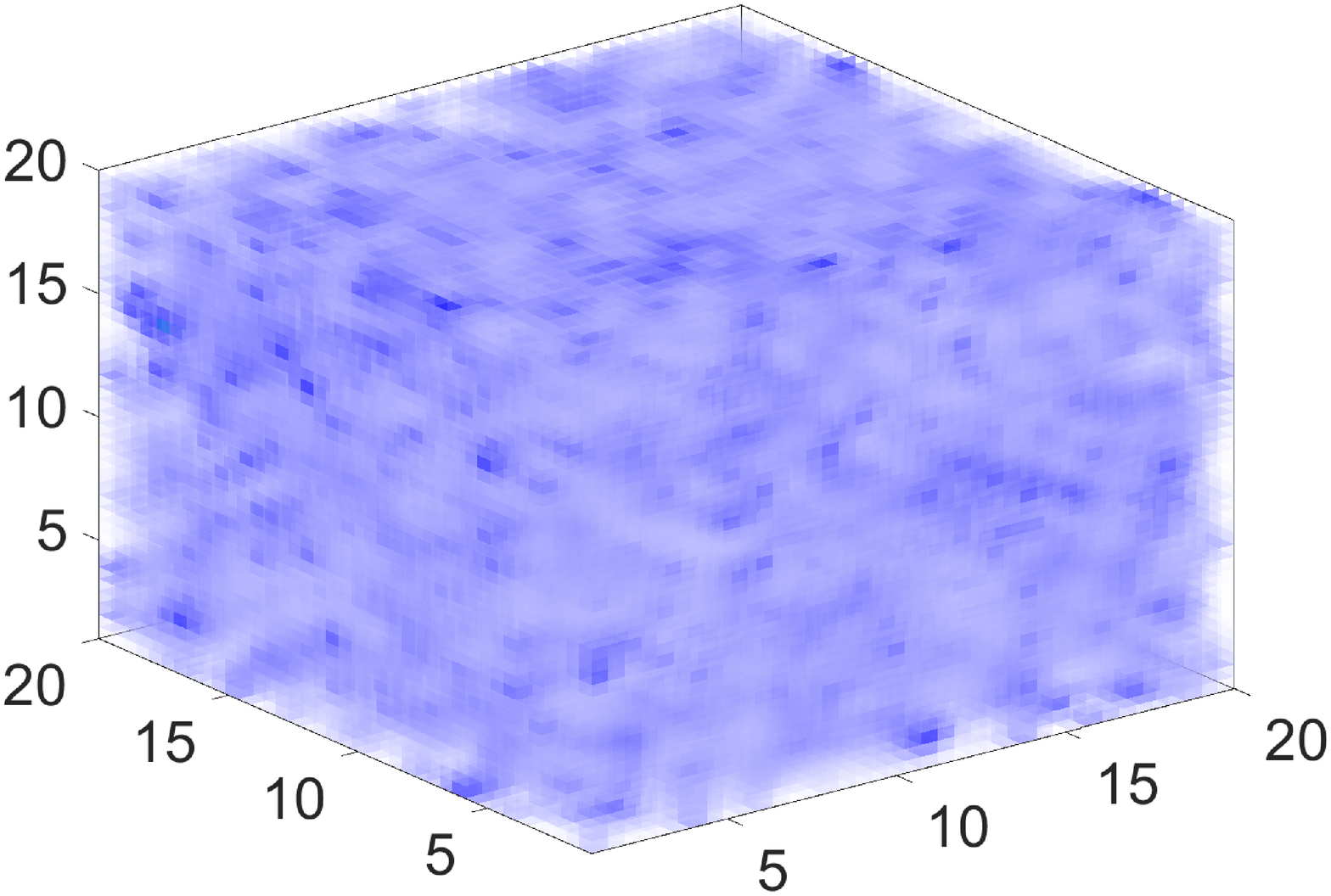}}
		\subfigure[]{\label{fig.spatial_2} 
			\includegraphics[width=8.5cm]{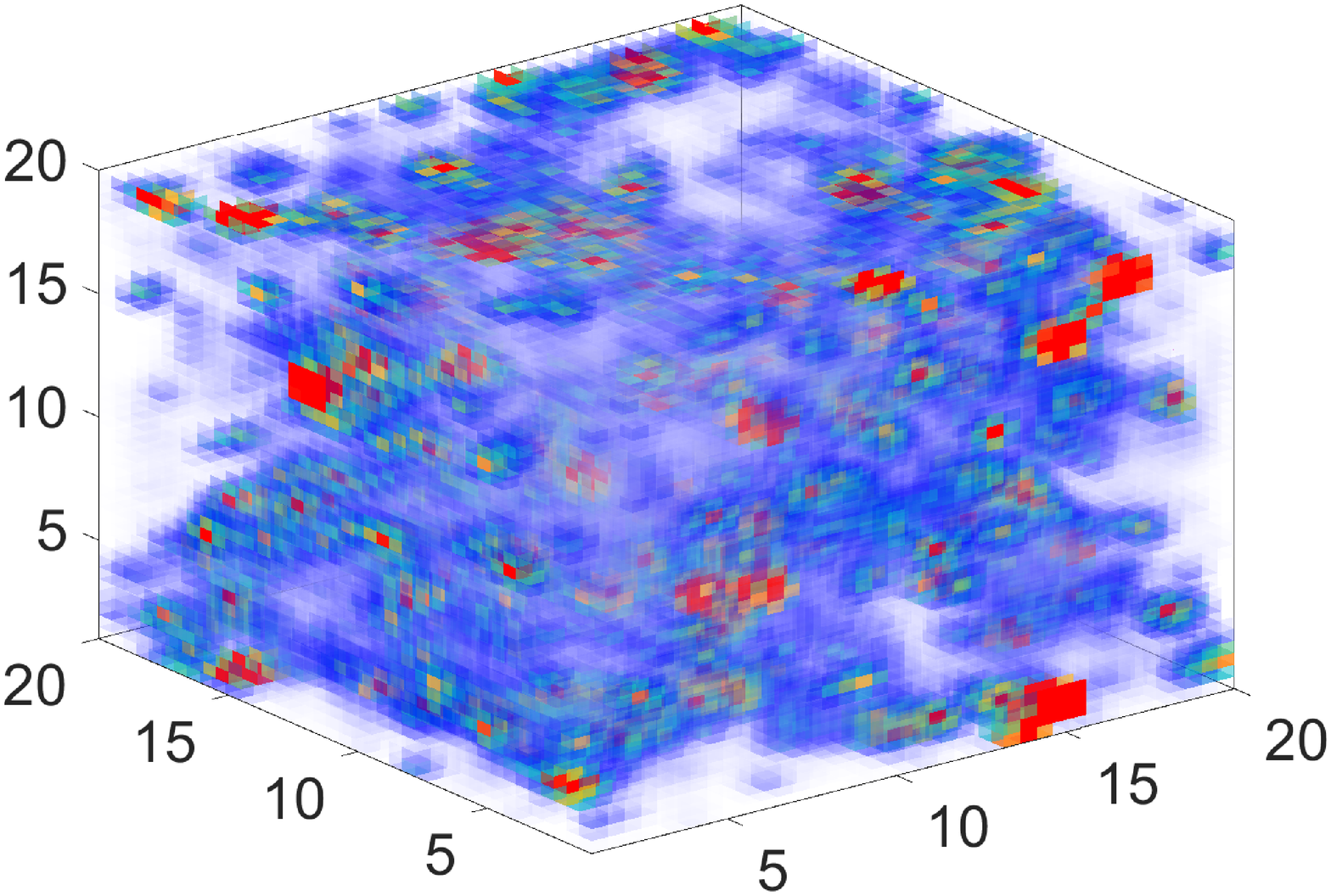}}
		\subfigure[]{\label{fig.spatial_3} 
			\includegraphics[width=8.5cm]{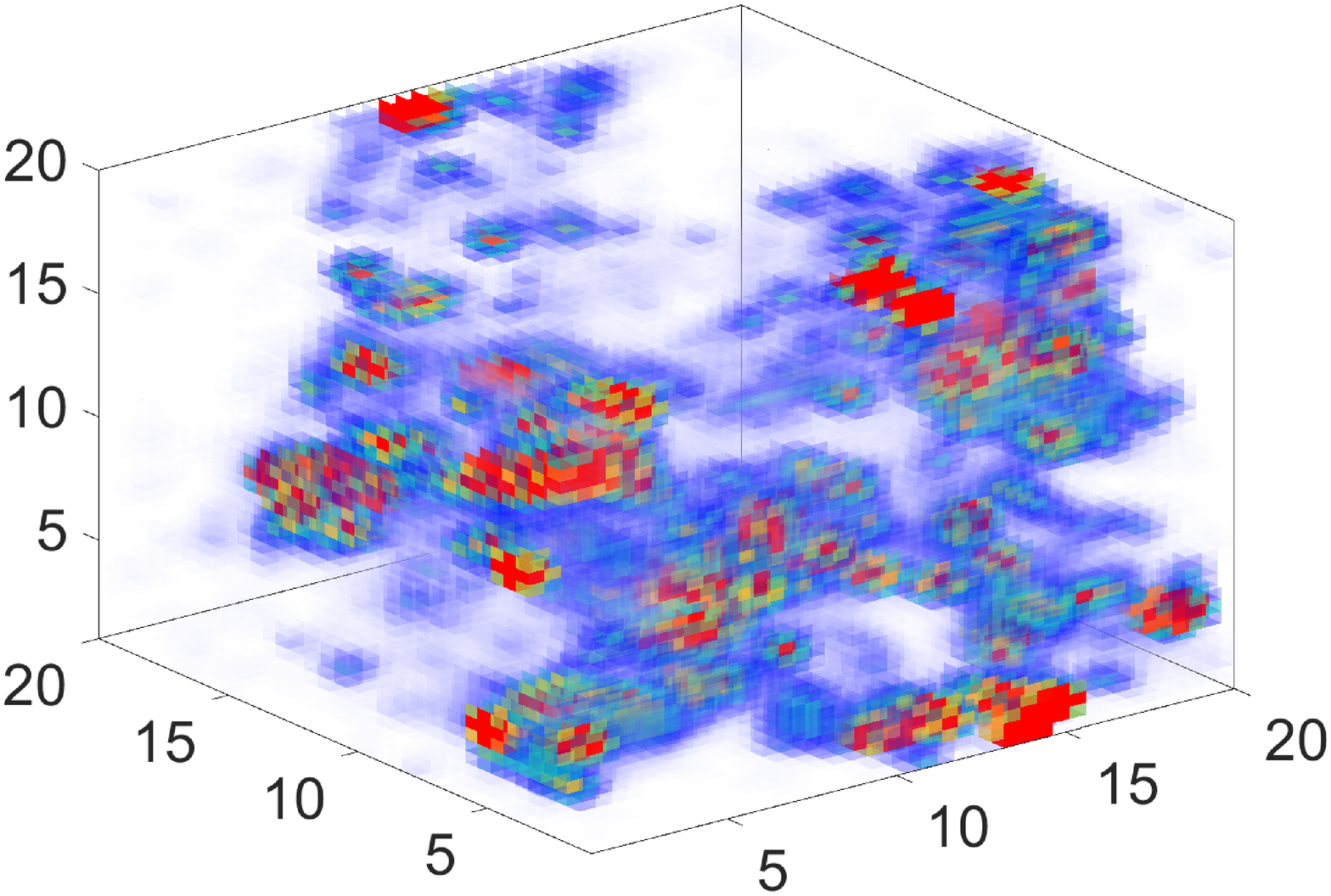}}
		\subfigure[]{\label{fig.spatial_4} 
			\includegraphics[width=8.5cm]{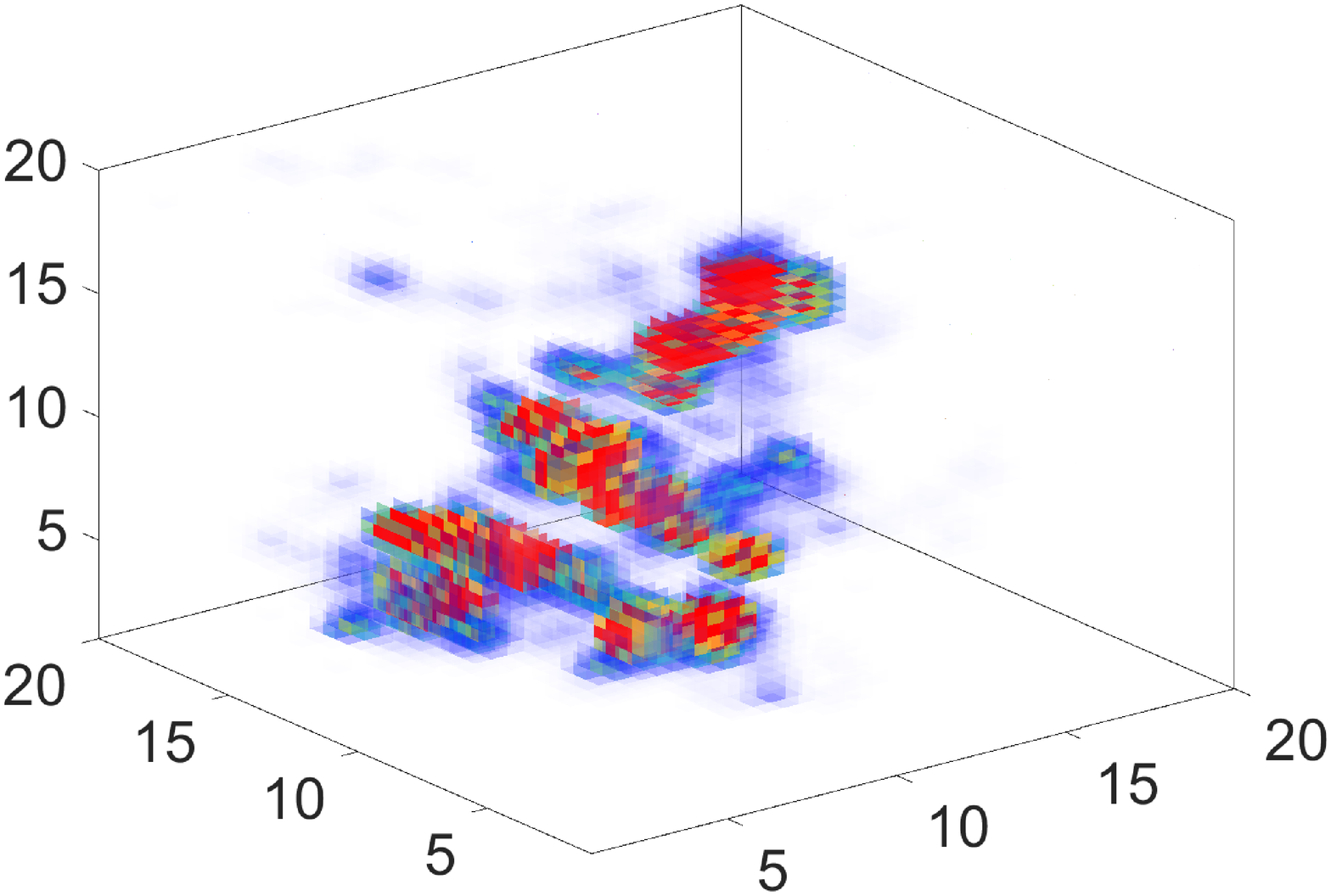}}
		\subfigure{}{\label{fig.spatial_5} 
			\includegraphics[width=8.5cm]{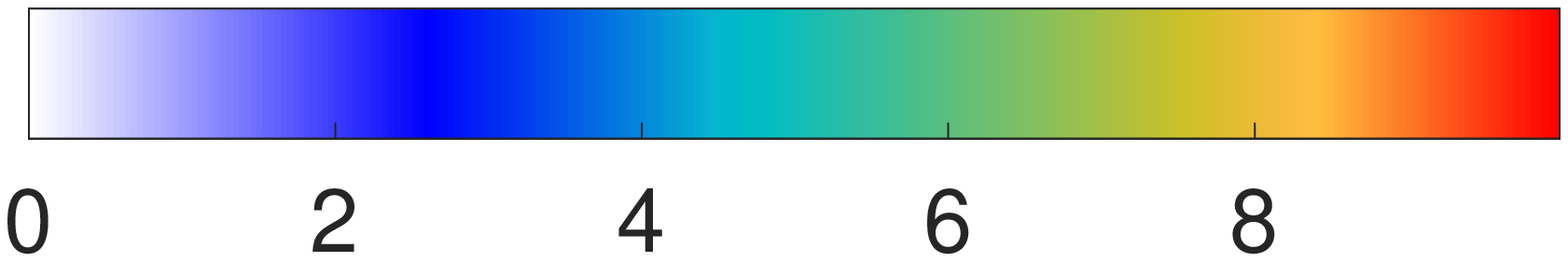}}
		\caption{The spatial distribution of the order parameter $\Delta(r_i)$ for a $20\times20\times20$ lattice. The cut-off energy $\omega_D = 2$, coupling constant $U = -1$ (both in units of $t$), and the density $\langle n \rangle = 0.875$. The disorder strength is $V = 4,10,12$ and $16$ from \subref{fig.spatial_1} to \subref{fig.spatial_4}. The order parameter amplitude $\Delta(r_i)$ is normalized by $\Delta_0 \sim 0.002$. As was expected, spatial inhomogeneities increase strongly with disorder. Especially for $V = 12$, we observe a rather intricate spatial pattern with large regions with an almost vanishing order parameter combined with localized splash corresponding to large enhancement of superconductivity that occur across the sample.}\label{Fig.spatial}
	\end{center}
\end{figure}

The spatial dependence of the order parameter $\Delta(r_i)$, resulting from the numerical solution of the BdG equations for a single disorder realization, is depicted in Fig.~\ref{Fig.spatial}. As was expected, $\Delta(r_i)$ becomes more spatially inhomogeneous as the strength of the random potential $V$ increases. For $V > 12$, is already rather localized in small regions of the sample which is an early indication that the transition could be located around that disorder strength. When $V=16$, the order parameter is concentrated in a small spatial region, which suggests that the transition to the insulating region has already taken place.

\begin{figure}
	\begin{center}
		\subfigure[]{\label{fig.gapdistribution_1}
			\begin{minipage}{0.485\linewidth}
				\includegraphics[width=8cm]{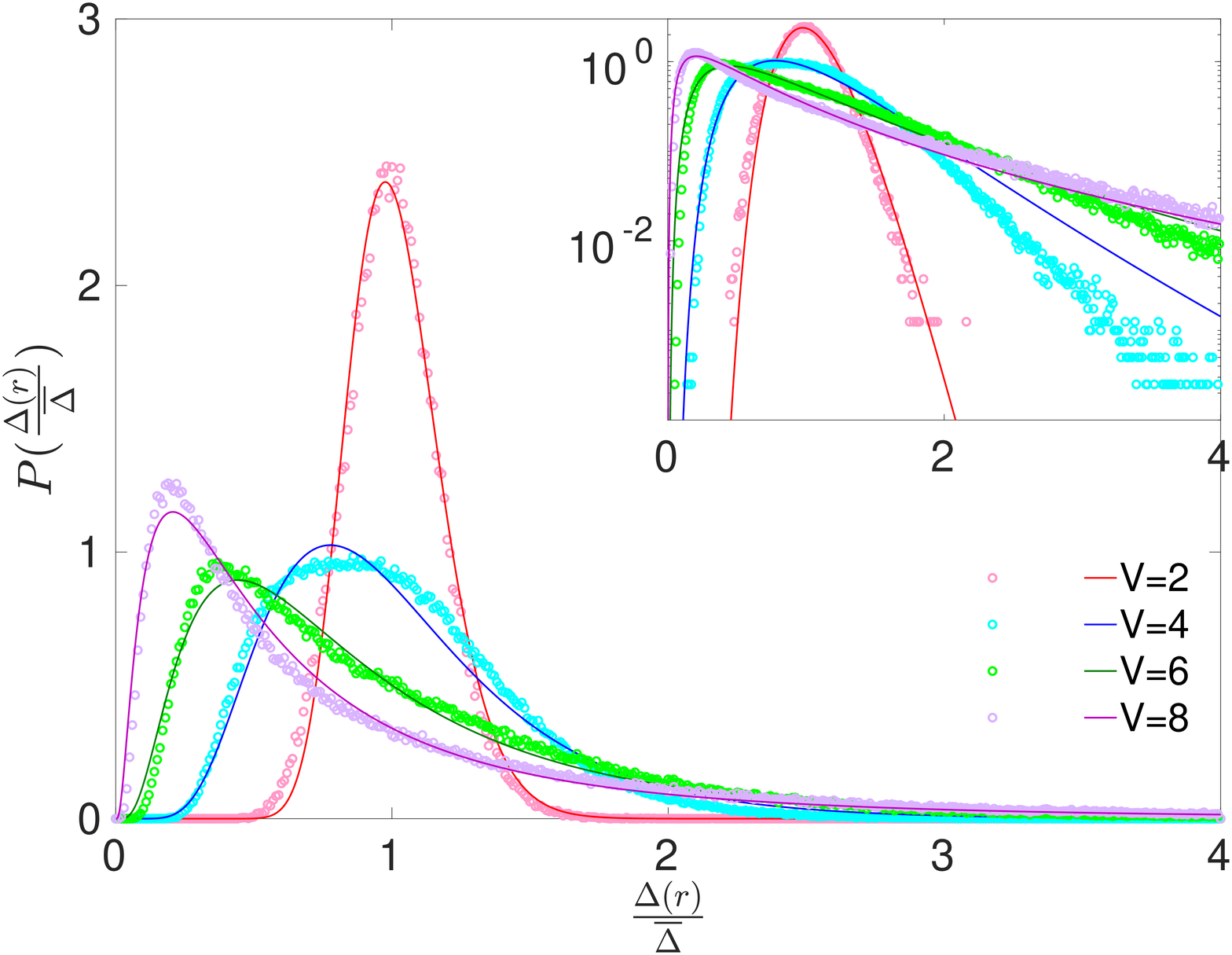}\\
				\includegraphics[width=8cm]{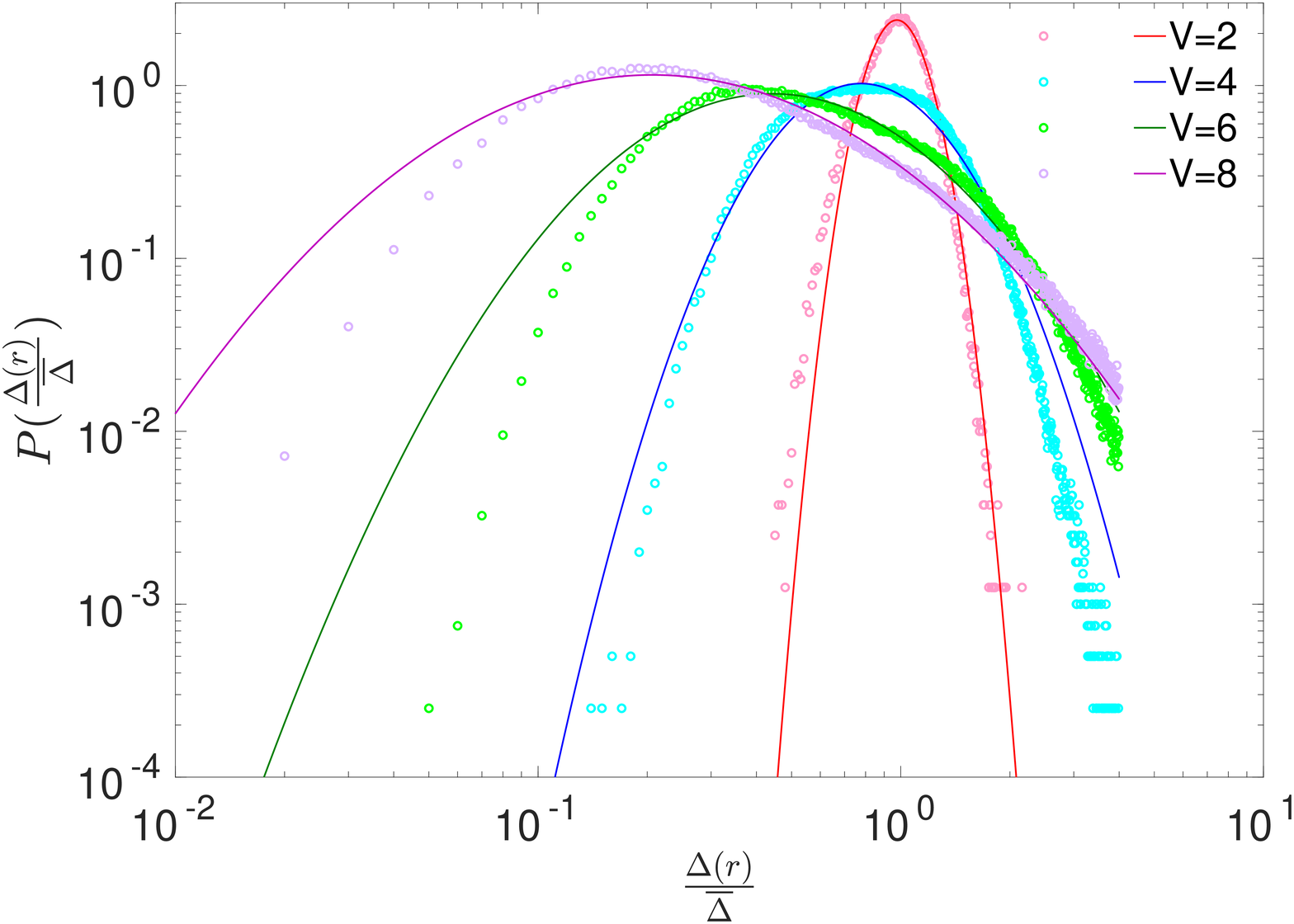}
				\vspace{0.05cm}
			\end{minipage}	
		}
		\subfigure[]{\label{fig.gapdistribution_2}
			\begin{minipage}{0.485\linewidth}
				\includegraphics[width=8cm]{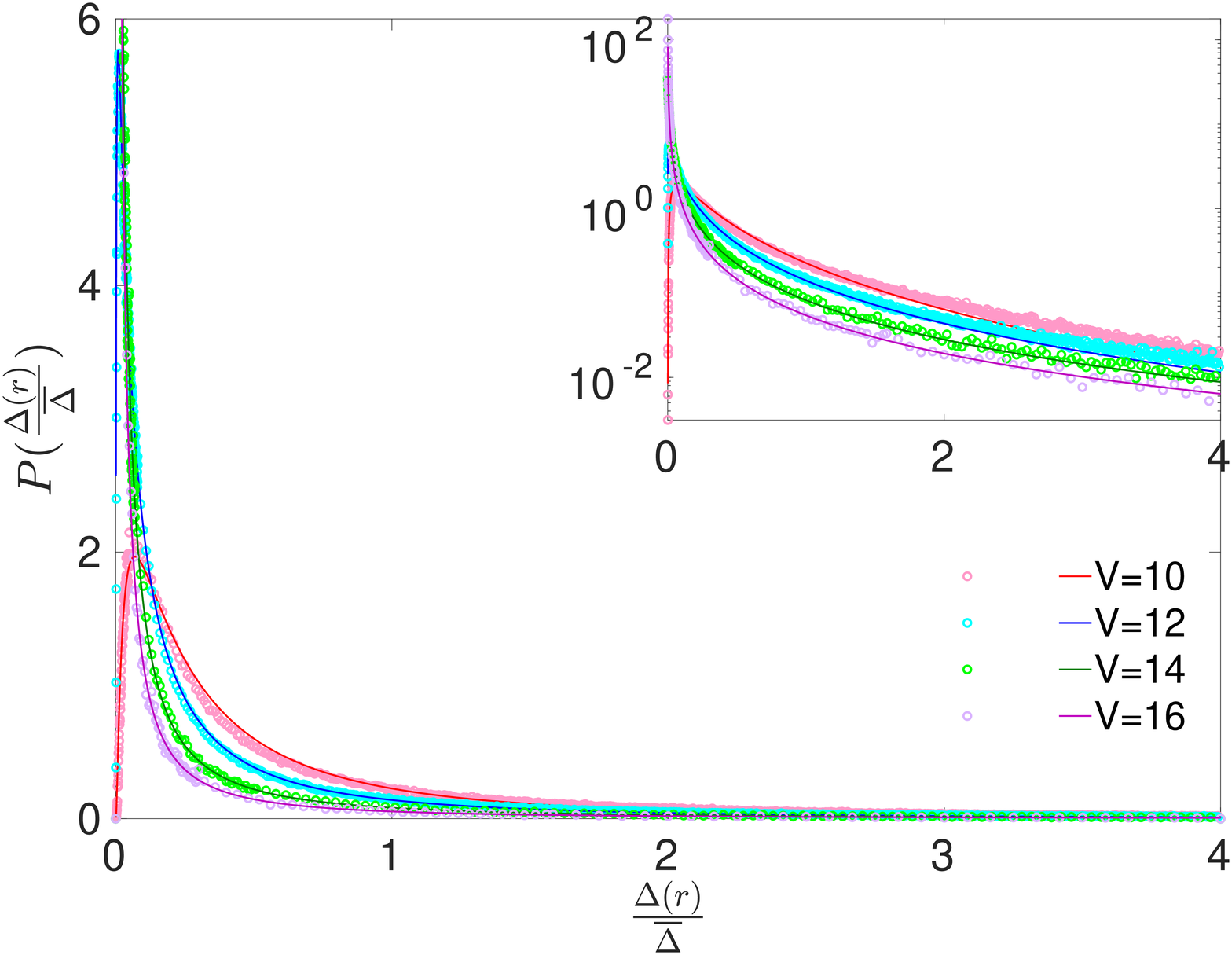}\\
				\includegraphics[width=8cm]{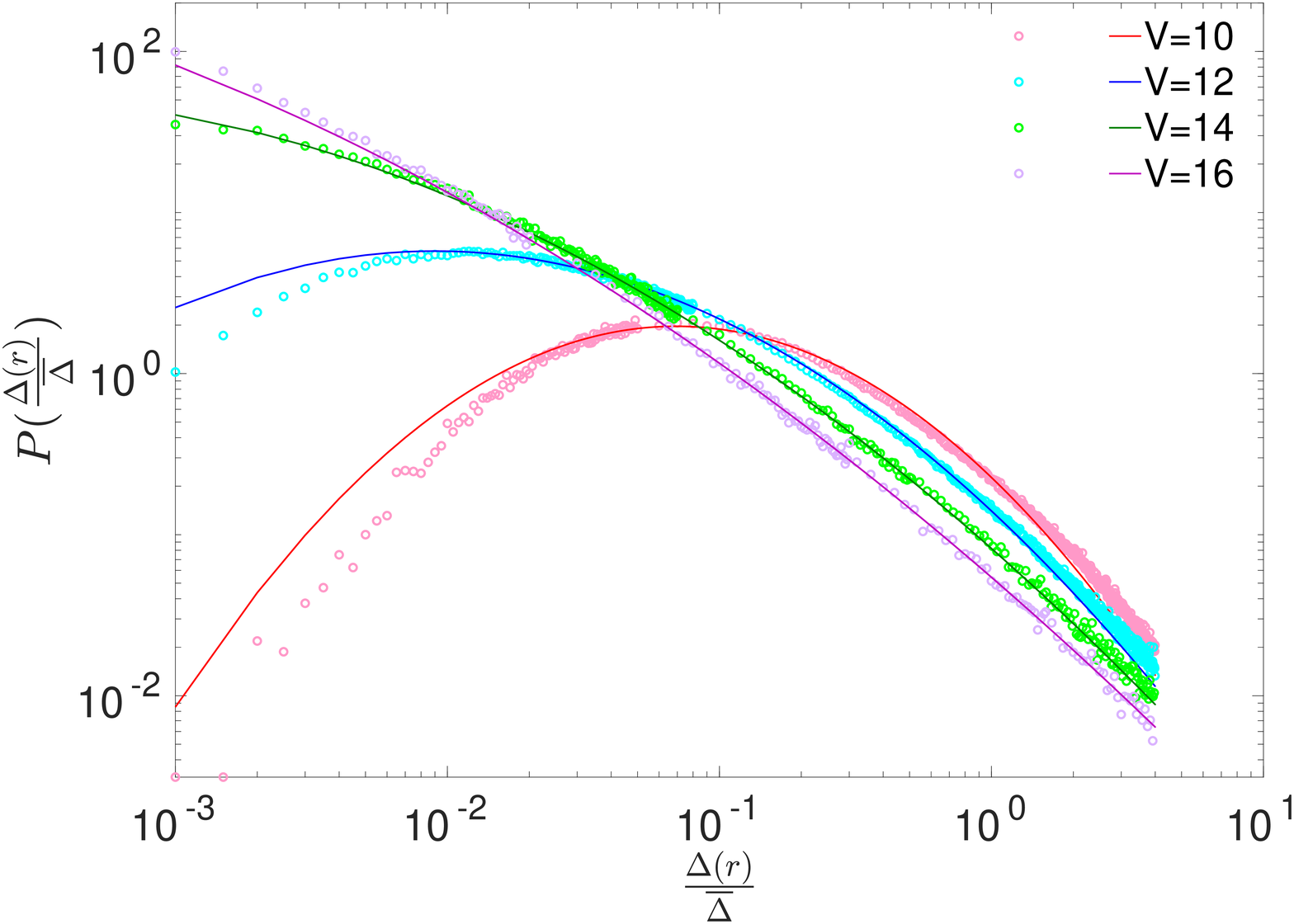}
				\vspace{0.05cm}
			\end{minipage}
		}
		
		\caption{The probability distribution of the order parameter $\Delta(r_i)$ (normalized by its spatial average $\bar{\Delta} \equiv \langle \Delta(r) \rangle$) for different disorder strength $V$. The numerical results (circle) are fit with a log-normal distribution Eq.~(\ref{eq.6}) (solid line). For weak disorder $V = 2$, the distribution is symmetric, relatively narrow and close to Gaussian. For intermediate disorder $V = 4 \sim 8$, it becomes broader, asymmetric and with an exponential tail. As disorder strength approaches the critical region, $V \sim 10$, the fitting to a log-normal distribution becomes increasingly accurate though with a maximum very close to zero $V \sim 12$ which indicates a very asymmetric distribution. }\label{Fig.gapdistribution}
	\end{center}
\end{figure}

The probability distribution of $\Delta(r_i)$, depicted in Fig.~\ref{Fig.gapdistribution}, captures accurately the gradual increase of spatial inhomogeneities. In the weak disorder region, the distribution is narrow and symmetric with a peak around the average order parameter. Deviations from a Gaussian distribution are small. As disorder increases, but still far from the transition, the distribution becomes broader and asymmetric. For $V \sim 6$, the tail of the distribution is well described by an exponential decay and, though asymmetric, the distribution has a clear maximum.

As disorder is further increased $V \geq 8$, the distribution becomes broader with tails that decay more slowly. We recall that, assuming that  eigenfunction correlations in the non-interacting limit are multifractal, it was found \cite{Mayoh2015} that the probability distribution for the order parameter $\Delta(r_i)$ of a two dimensional superconductor in the weak-coupling, weak-disorder limit is log-normal,
\begin{equation}
P\left(\frac{\Delta(r)}{\bar{\Delta}}\right) = \frac{\bar{\Delta}}{\Delta(r)\sqrt{2\pi}\zeta}\exp\left(-\frac{\left[\ln\left(\frac{\Delta(r)}{\bar{\Delta}}\right)-\eta\right]^2}{2\zeta^2}\right),	\label{eq.6}
\end{equation}
where $\zeta$ and $\eta$ are disorder dependent constants.
Surprisingly, we find an increasingly good agreement with the log-normal distribution. The singularity spectrum, depicted in Fig.~\ref{Fig.falpha}, is still parabolic in this range of parameters. This parabolicity is directly related to the spectrum of multifractal dimensions that enters in the analytical derivation \cite{Mayoh2015} of the probability distribution in the 2D case. Indeed, in Ref.~\cite{Feigelman2010}, the analytical calculation of the moments of the order parameter at the 3D Anderson transition were consistent with this result.

As disorder further increases, when $V \geq 10$, the maximum of the distribution shifts to small values of the order parameter. The tail becomes broader with an even slower decay. Overall, the distribution is still well described by a log-normal distribution. 

As can be observed in Fig.~\ref{Fig.gapdistribution} for $V \approx 12$, the maximum is not noticeable and the distribution is flat for very small values of the order parameter. This indicates that in a substantial number of points, the order parameter either vanishes or is much smaller than the bulk value for no disorder. We find it plausible that the insulating transition occurs precisely at this disorder strength.
For stronger disorder, corresponding to the insulating region, the decay seems to become power-law. This regime will be discussed in more detail in a forthcoming publication \cite{us2020}.

\subsection{Singularity spectrum of the order parameter amplitude distribution}
In order to obtain further information about the spatial distribution of the order parameter, we now compute the singularity spectrum $f(\alpha)$ \cite{Jensen1989}. More specifically, we aim to clarify to what extent the order parameter amplitude inherits the multifractality \cite{Wegner1980,Castellani1986} of eigenstates observed in the non-interacting limit, and approximately, for what disorder strength, the superconductor-insulator transition occurs. 

\begin{figure}
	\begin{center}
		\subfigure[]{\label{fig.falpha_1} 
			\includegraphics[width=8.5cm]{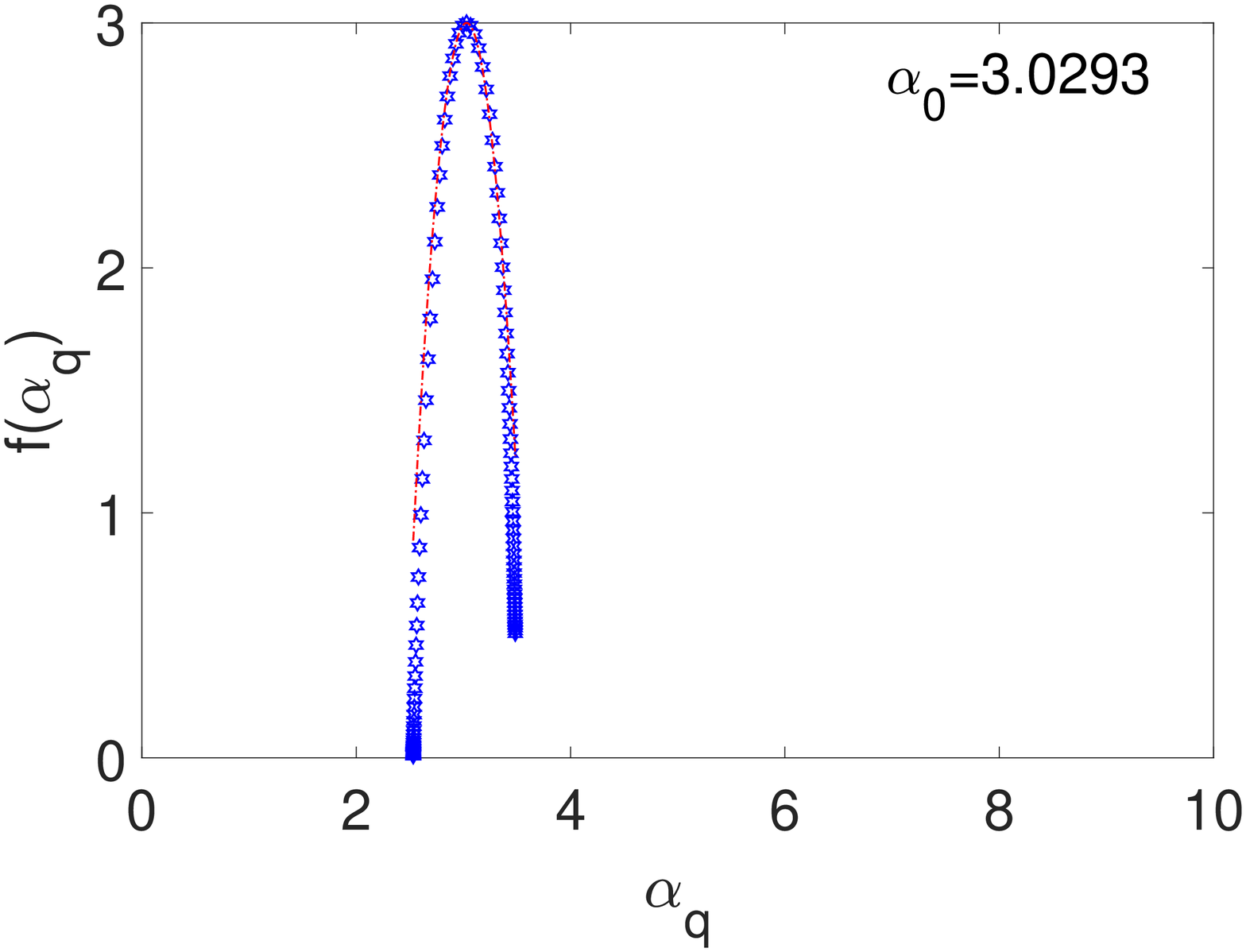}}
		\subfigure[]{\label{fig.falpha_2} 
			\includegraphics[width=8.5cm]{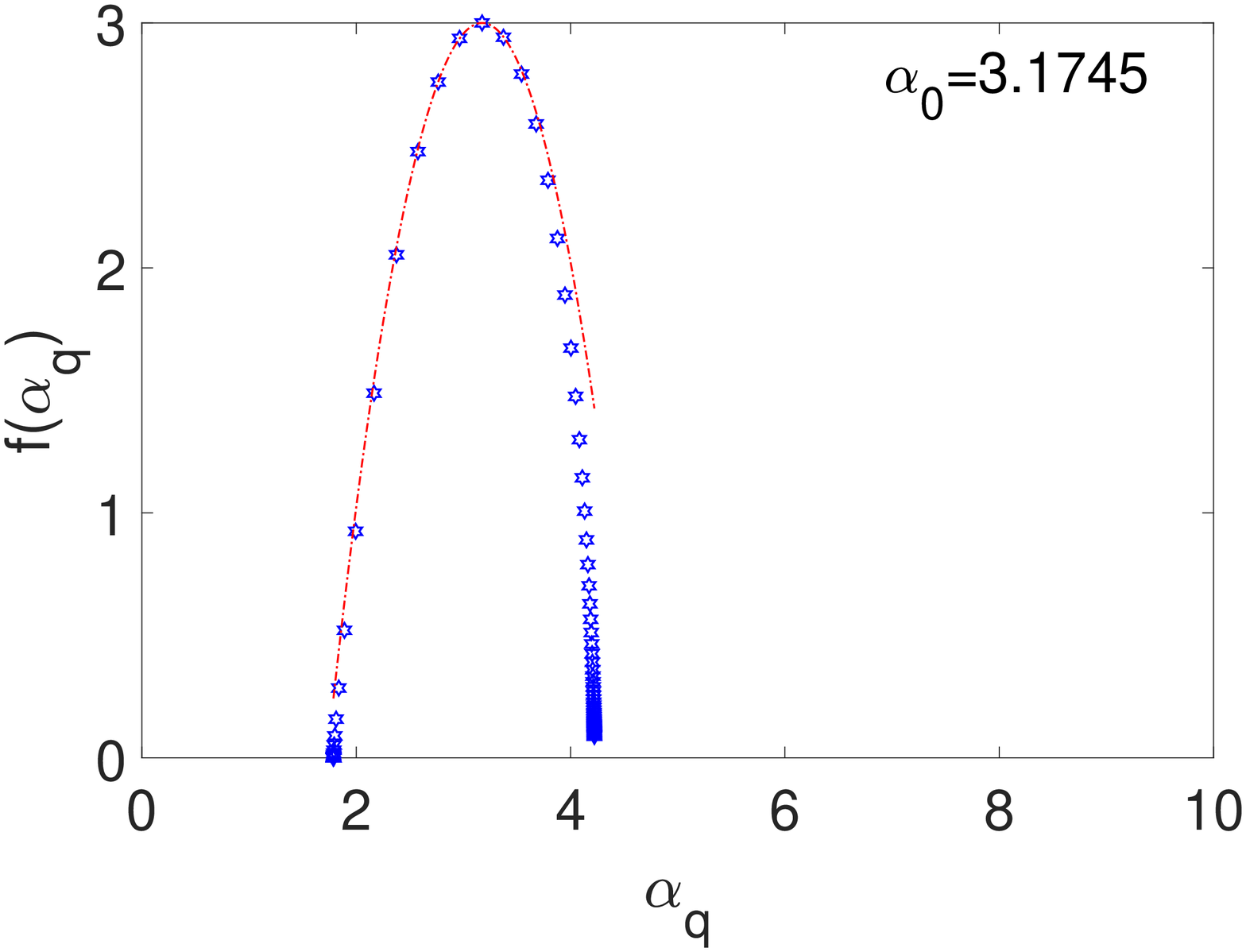}}
		\subfigure[]{\label{fig.falpha_3} 
			\includegraphics[width=8.5cm]{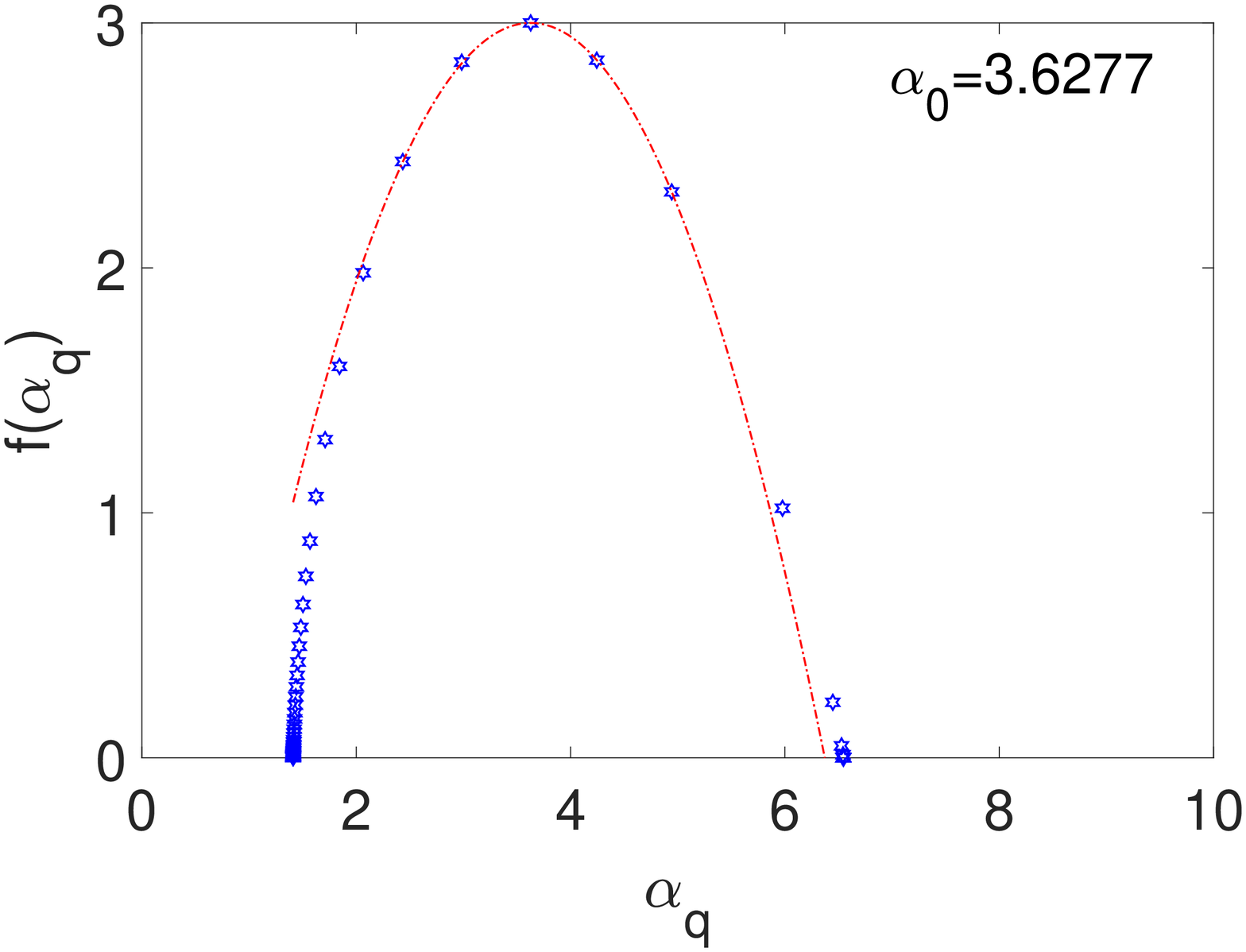}}
		\subfigure[]{\label{fig.falpha_4} 
			\includegraphics[width=8.5cm]{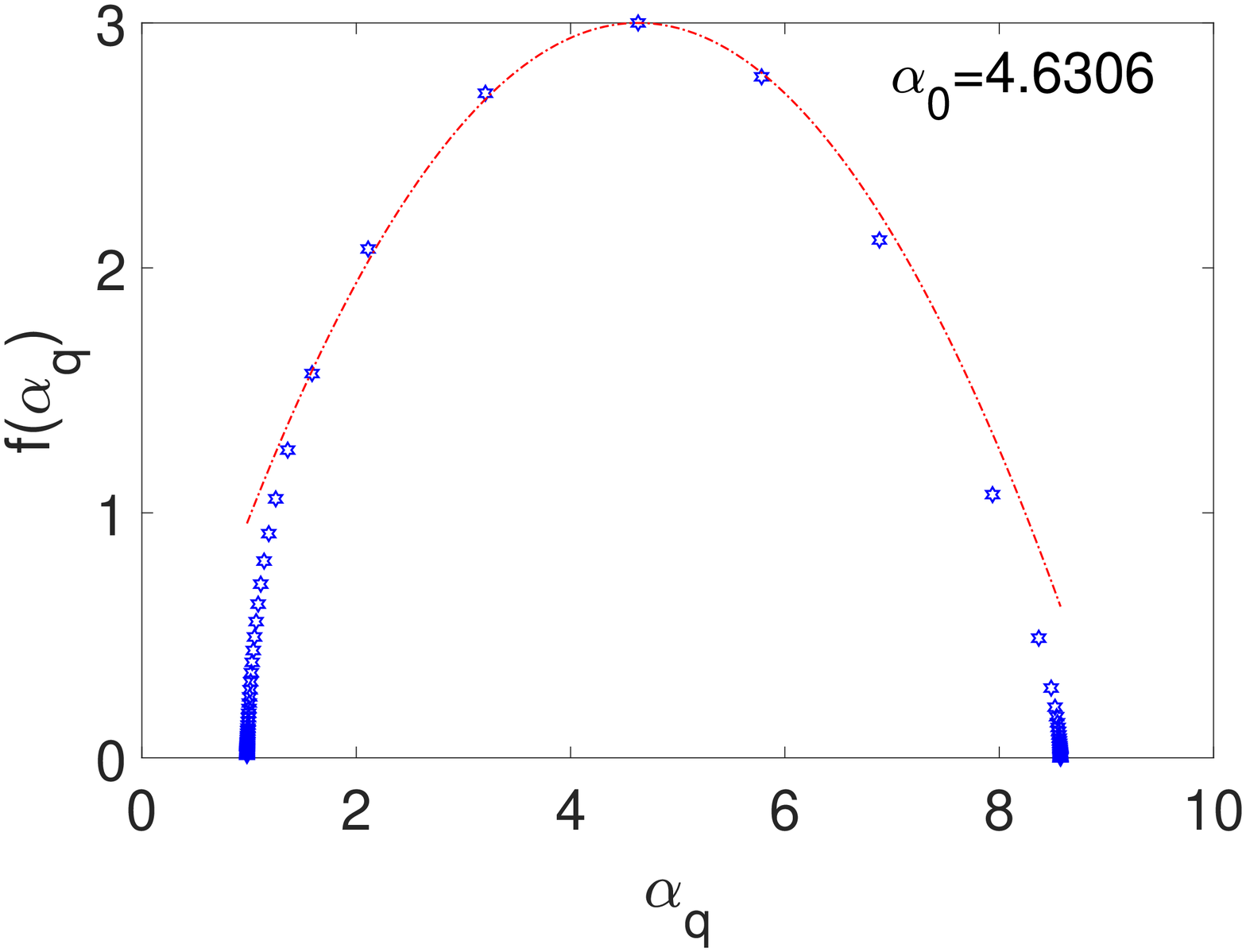}}
		\caption{The singularity spectrum $f(\alpha)$ related to the order parameter $\Delta(r_i)$ for a $20\times20\times20$ lattice size and for $V = 4,8,12$ and $16$ from \subref{fig.falpha_1} to \subref{fig.falpha_4}. The cut-off energy $\omega_D = 2$, coupling constant $U = -1$, and the density $\langle n \rangle = 0.875$. It agrees well with the parabolic prediction (dotted line) corresponding to multifractal eigenstates. Also in agreement with the theoretical prediction, the parabolic curve becomes broader and its maximum shifts to larger values as disorder increases. The only exception is \subref{fig.falpha_4}, for $V=16$, which is in the insulator region. The parabolic fitting only describes well around the central part of the singularity spectrum but not the observed termination of multifractal dimensions. This is an indication, together with the large value of $\alpha_0 = 4.6306$, that the system is no longer critical at this disorder strength.}\label{Fig.falpha}
	\end{center}
\end{figure}

In the non-interacting limit, the singularity spectrum, also called $f(\alpha)$ spectrum, is related to the scalings of the density of probability associated to multifractal eigenstates at the Anderson transition. In 2D, eigenstates are approximately multifractal for weak disorder provided that system size is much smaller than the localization length. In this weak multifractal region, the $f(\alpha)$ spectrum is parabolic \cite{Wegner1980}. A qualitatively similar parabolic singularity spectrum \cite{mildenberger2002,Evers2000,Evers2008} is a feature of the 3D Anderson transition despite the fact that the transition occurs at strong disorder. 

From Eq.~\eqref{eq.3}, $\Delta(r_i)$ is given by a self-consistent condition, which is a weighted average over the eigenstates $u_n(r_i)$ and $v_n(r_i)$ of the BdG equations. At least for clean nano-grains \cite{Shanenko2007}, it was found that $u_n(r_i)$ and $v_n(r_i)$ are proportional to the eigenstates of the one-body problem $\Psi_n(r_i)$ for sufficiently weak coupling. Therefore, it seems plausible, especially if the weighted sum defining $\Delta(r_i)$ does not contain many eigenstates, that some of the anomalous scaling features, reflected in the singularity spectrum of the eigenstates of the one-body problem, may be inherited by the order parameter.

\begin{figure}
	\begin{center}
		\subfigure[]{\label{fig.alpha0_1} 
			\includegraphics[width=10cm]{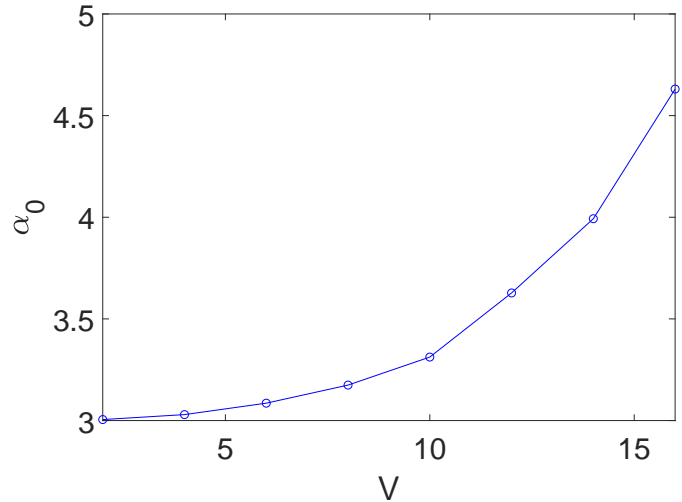}}
		\caption{$\alpha_0$ as a function of disorder. For relatively weak disorder $V\leq10$, $\alpha_0$ changes slowly with disorder. However in the critical region, $V \sim 12$, the increase is faster, which suggests a stronger spatial inhomogeneity.}\label{Fig.alpha0}	
	\end{center}
\end{figure}

In order to carry out the computation, we define $|P(r_i)|^2 = \frac{\Delta(r_i)}{\sum_{j=1}\Delta(r_j)}$ and compute the $f(\alpha)$ spectrum of $|P(r_i)|^2$ following the method introduced in Ref.~\cite{Jensen1989}. The results for disorder strengths $V=4,8,12$ and $16$ are depicted in Fig.~\ref{Fig.falpha}. We find that the singularity spectrum $f(\alpha)$ for intermediate disorder $V \sim 12$ is well approximated by $f(\alpha) = 3 - \frac{(\alpha - \alpha_0)^2}{4(\alpha_0 - 3)}$, with $\alpha_0 \sim 4$. Approximately, this is the analytical prediction \cite{Evers2008,mildenberger2002} for the three dimensional system at the Anderson transition. Moreover, precisely in this region, the parameter $\alpha_0$, depicted in Fig.~\ref{Fig.alpha0}, that controls the broadness of the singularity spectrum, experiences a faster increase with disorder.
These results point to a spatial distribution of the order parameter characterized by multifractal-like spatial structure. We will confirm this prediction in section \ref{sec:levelstatistics} by a detailed analysis of the level statistics of the system.

We note that for $V = 16$, clear deviations from a parabolic spectrum are observed and the fitted $\alpha_0$ is larger than the prediction for the Anderson transition in three dimensional non-interacting systems. This suggests that the system is already an insulator and that therefore the critical disorder at which the transition occurs is around $V \sim 12$. 

Having shown that at certain disorder strength, 
the order parameter may have multifractal features. We study in next section how many eigenstates contribute effectively to the formation of the order parameter, especially around this critical region. This is important as the level statistic analysis must be restricted to the spectral window relevant for the formation of the Cooper pairs.

\section{What eigenstates $u_n(r)$ and $v_n(r)$ contribute to $\Delta(r)$?}\label{sec:overlap}
In order to have a more quantitative understanding about how exactly $\Delta(r)$ is built up from the eigenfunctions $\{u_n(r),v_n(r)\}$ of the BdG equation, we study, 
\begin{equation} 
P_{uv} = \sum_{r}|u_n^2(r)-v_n^2(r)|. \label{puv}
\end{equation}
A strong overlap of $u_n$ and $v_n$ corresponds to $P_{uv} \approx 0$, while if $u_n$ and $v_n$ are completely decoupled, then $P_{uv} \approx 1$ since $\sum_{r}(u_n^2(r)+v_n^2(r)) = 1$. We note that, because of the self-consistent condition Eq.~\eqref{eq.3}, only eigenstates $u_n$ and $v_n$ that overlap strongly contribute significantly to $\Delta(r)$. Therefore, the study of $P_{uv}$ will reveal how many eigenstates effectively contribute to the formation of the order parameter. This will be important later for the determination of of the critical disorder at which the transition to localization occurs.

\begin{figure}
	\begin{center}
		\subfigure[]{\label{fig.puv_1} 
			\includegraphics[width=8.5cm]{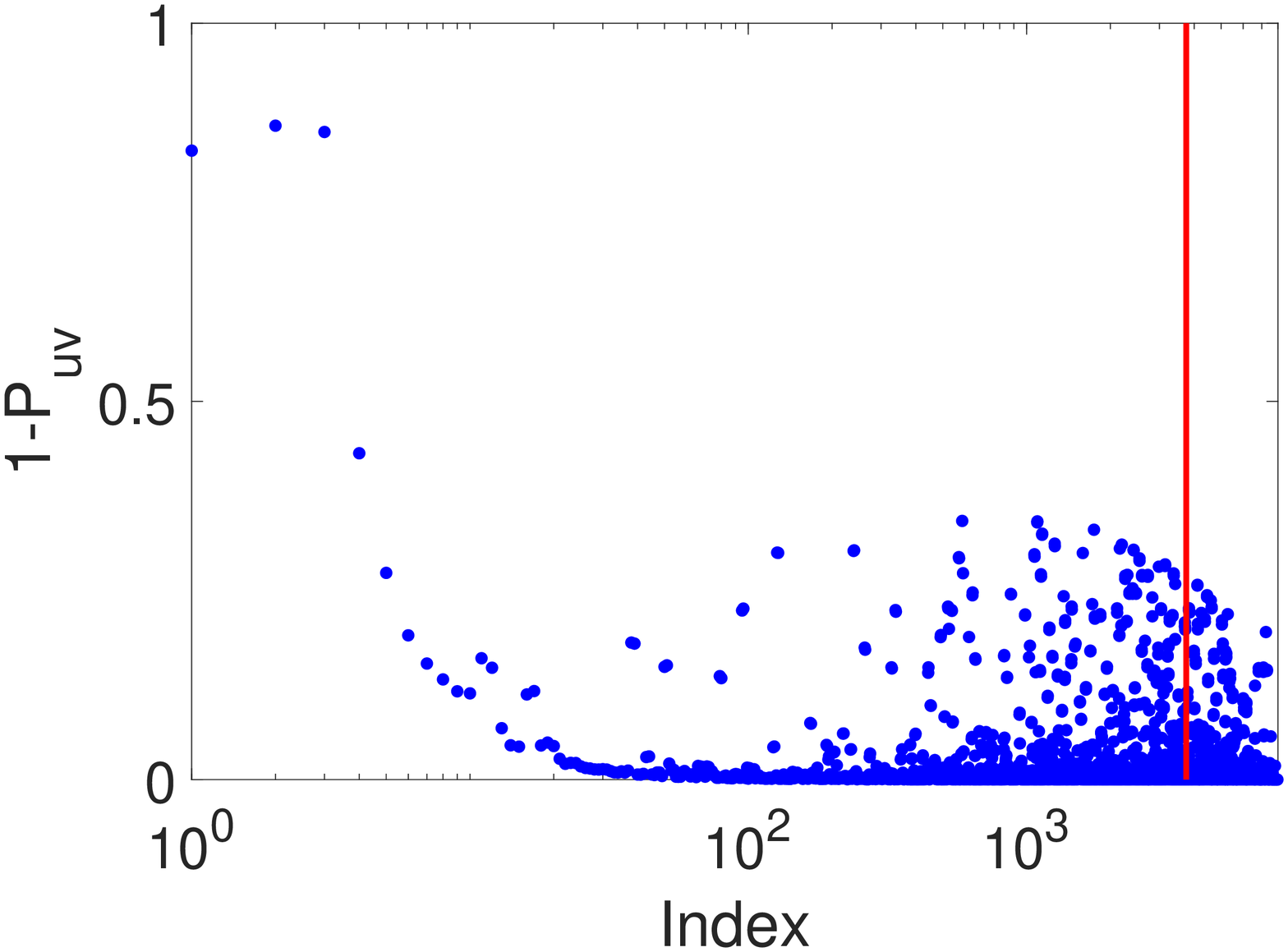}}
		\subfigure[]{\label{fig.puv_2} 
			\includegraphics[width=8.5cm]{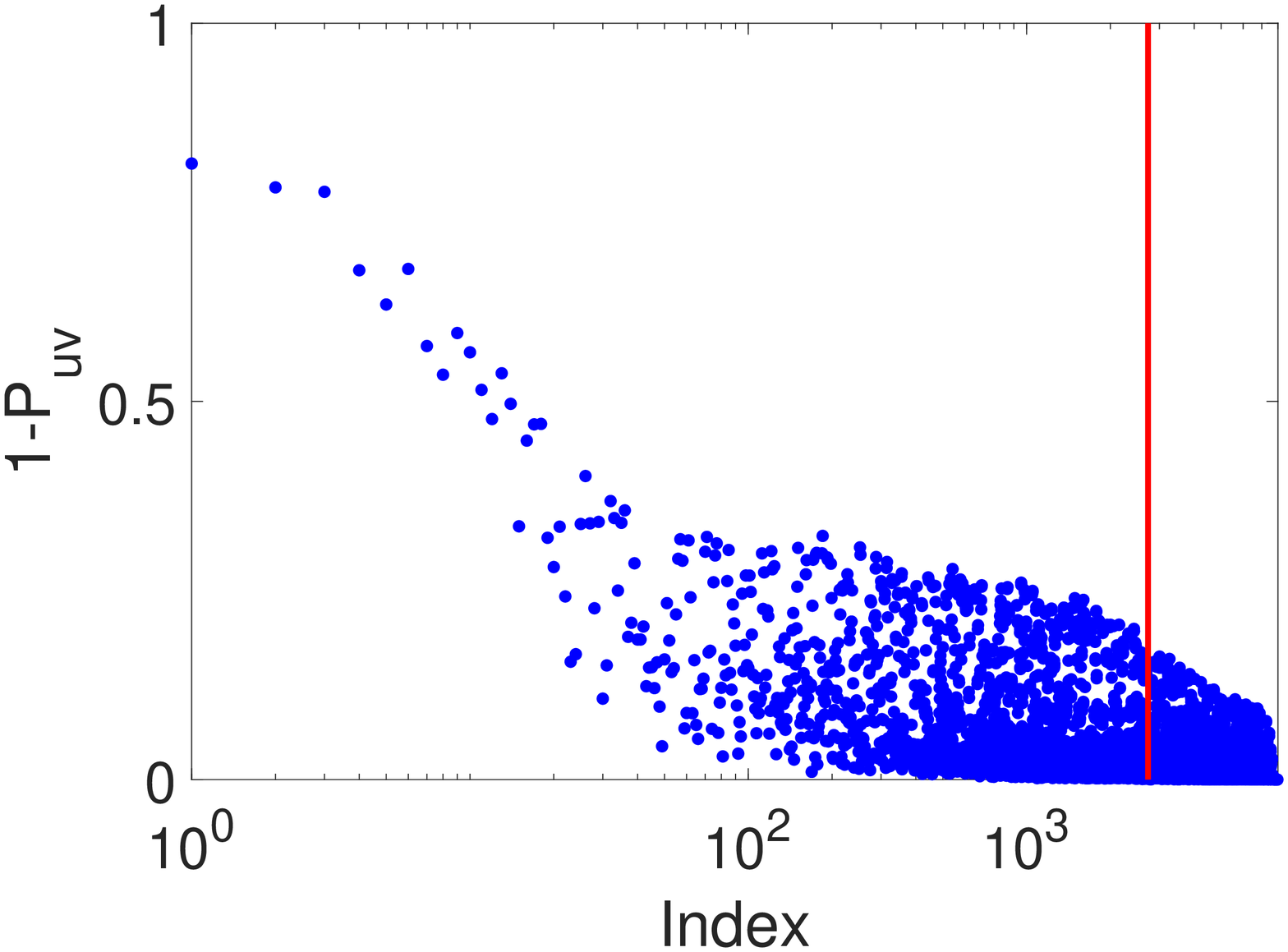}}\\
		\subfigure[]{\label{fig.puv_3} 
			\includegraphics[width=8.5cm]{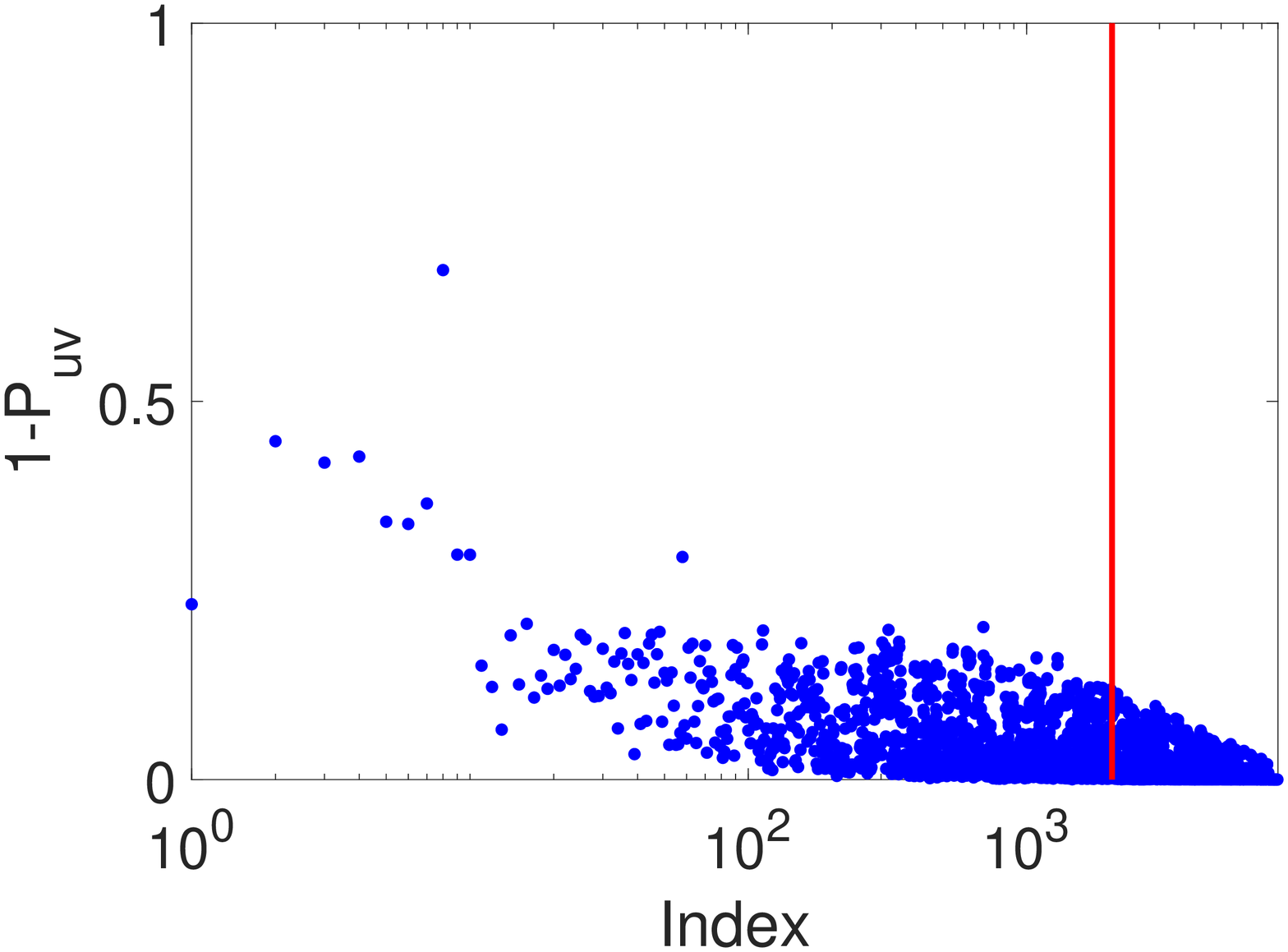}}
		\subfigure[]{\label{fig.puv_4} 
			\includegraphics[width=8.5cm]{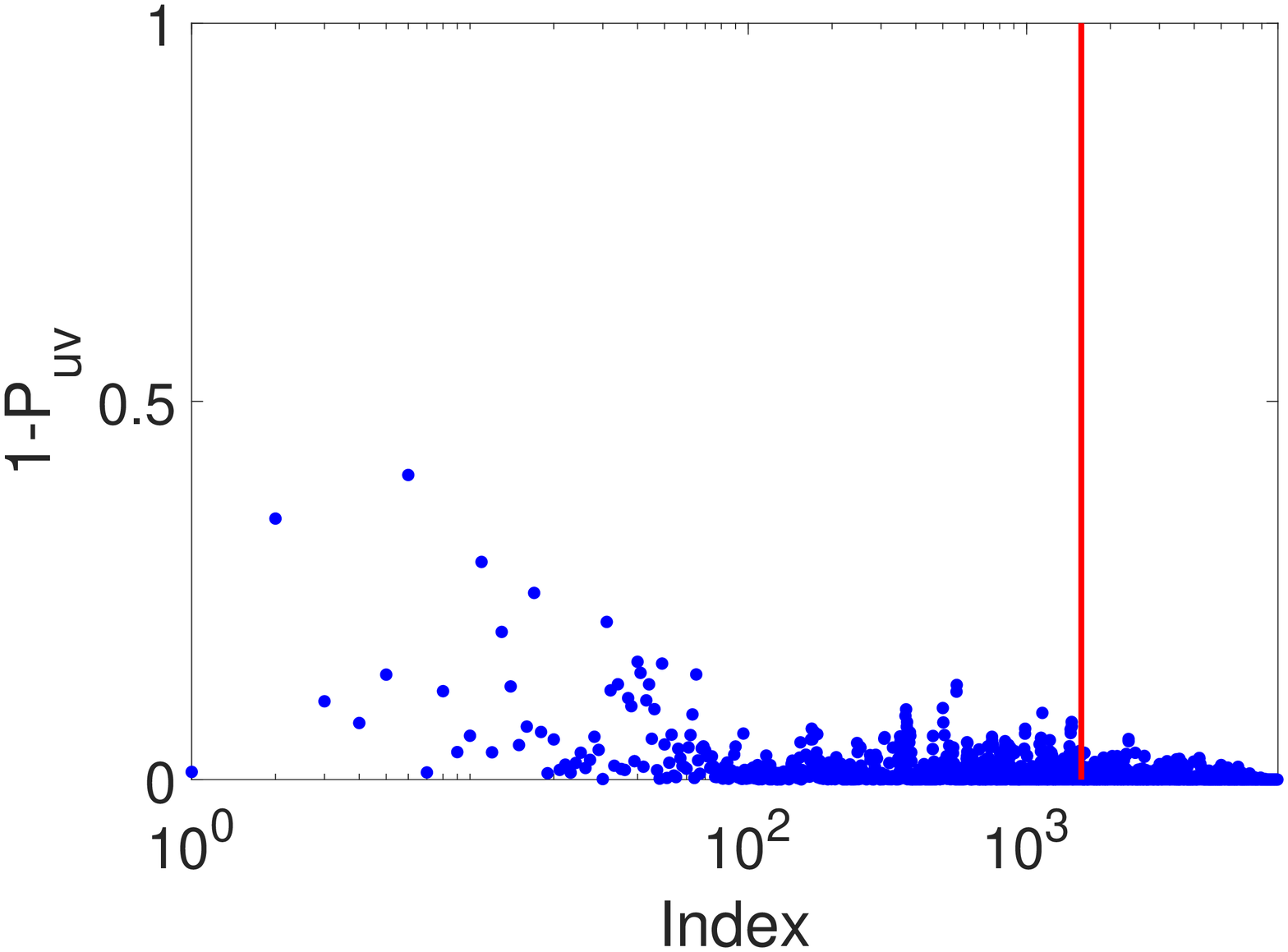}}\\
		\caption{The coupling between $u_n$ and $v_n$. $1-P_{uv}$, as expressed in Eq.~\eqref{puv}, for different disorder strength $V =4, 8, 12$, and $16$ from \subref{fig.puv_1} to \subref{fig.puv_4}. The vertical red line shows the position of the cut-off energy $\omega_D=2$. For disorder $V\leq 10$, eigenfunctions corresponding to the lowest eigenvalues, are almost identical and therefore $1-P_{uv} \approx 1$. However, for disorder $V = 12$, $1-P_{uv} \approx 0.5$ even for the lowest  eigenfunctions. In the insulator region $V=16$,  eigenfunctions are localized, which results in a weak overlap and therefore in a even smaller $1-P_{uv}$. For $V \leq 8$, the number of strongly correlated eigenstates increases with disorder, compare \subref{fig.puv_1} and \subref{fig.puv_2}, which explains why disorder enhances superconductivity, see Fig.~\ref{Fig.gap}.}\label{Fig.puv}
	\end{center}
\end{figure}
Results, depicted in Fig.~\ref{Fig.puv}, show that only for a small number of eigenstates near $E = 0$, which is much less than the total number of states contained in the Debye energy window, the overlap is strong so that $P_{uv}$ is close to $0$. For the rest, $P_{uv} \approx 1$ which strongly suggests that only a small set of eigenvectors participate in the construction of the order parameter $\Delta(r_i)$. 
Interestingly, as disorder increases, the number of strongly coupled eigenstates $P_{uv} \approx 0$ increases as well. However, for $V\geq 12$, it seems that the trend is reversed. Fewer eigenstates contribute, and the overlap strength is weaker. Even for eigenstates very close to $E = 0$, $P_{uv}$ is never close to zero.

Taking into account that, through the self-consistent condition Eq.~\eqref{eq.3}, $\Delta(r_i)$ is also directly related to the overlap between $u_n(r_i)$ and $v_n(r_i)$. It is not surprising that the spatial average of $\Delta(r_i)$ increases with $V$ up to $V \sim 10$ where the increase stops and finally decreases for stronger disorder.
Effectively, as disorder increases, more eigenstates contribute to the formation of the order parameter which, as we said, will likely help its enhancement. More quantitatively, as depicted in Fig.~\ref{fig.puv_2}, more than $100$ states are strongly coupled for $V = 8$. However, such strong correlation is restricted to no more than $20$ eigenvectors for $V = 4$, see Fig.~\ref{fig.puv_1}.

\begin{figure}
	\begin{center}
		\includegraphics[width=10cm]{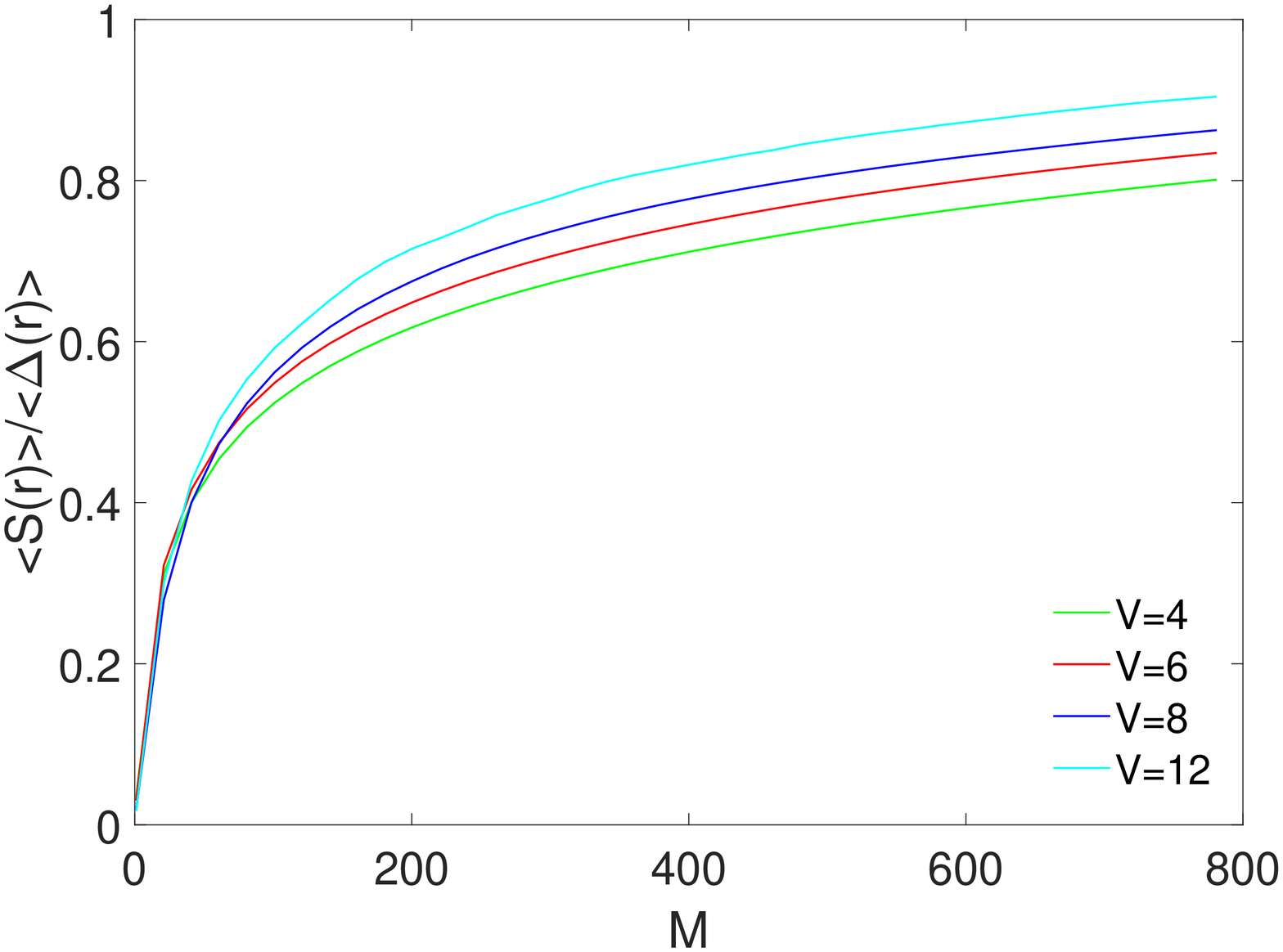}
		\caption{$\langle S(r) \rangle / \langle \Delta(r) \rangle$ as a function of $M$, the number of states, starting from the ground, that are taken into consideration to obtain $\langle S(r) \rangle$.  The system size is $20\times20\times20$, $\omega_D = 2$, $U = -1$, and the density $\langle n \rangle=0.875$. 
	About the first $100$ states, corresponding to $3\%$ of states inside the Debye window, contribute to more than the $50\%$ of the value of the order parameter. This percentage is larger as disorder increases.}\label{Fig.S_uv_s}
	\end{center}
\end{figure}

With the chosen Debye energy, about $35\%$ of eigenstates, around $3000$ states for size $20\times20\times20$, as is depicted in Fig.~\ref{fig.puv_1}, contribute to the order parameter. However, see Fig.~\ref{Fig.puv}, only a very small part of states near $E=0$ contributes significantly to the build up of the order parameter. For a more quantitative estimation, we define $S(r_i) = |U|\sum_{n=1}^{M}u_n(r_i)v_n^*(r_i)$ which for sufficiently large $M$ becomes the order parameter. We only show the first $800$ states in Fig.~\ref{Fig.S_uv_s}, which already represent more than $80\%$ of the total value of $\langle \Delta(r) \rangle$. 

More interesting is the fact that, only the first $100$ states, that represent about $3\%$ of the allowed eigenstates in the Debye window, are responsible for more than $50\%$ of the value of $\langle \Delta(r) \rangle$. 
Indeed, if we only take the first $10$ eigenstates into consideration, $\langle S(r) \rangle$ still reproduces a sizable part of $\langle \Delta(r) \rangle$, which is weakly dependent on the  considered disorder strength. 

These results are consistent with the overlap of eigenfunction $\{u_n, v_n\}$ shown in Fig.~\ref{Fig.puv}. About $100$ states closer to $E = 0$ are strongly coupled when $V=8$, while less than $20$ states are strongly coupled when $V=4$. Moreover, the coupling of $u(r)$ and $v(r)$ for the first $10$ eigenstates, is qualitatively similar for the different disorder strength, which results in a similar $\langle S(r) \rangle/\langle \Delta(r) \rangle$ in this region. Therefore, a relatively small number of strongly coupled eigenstates close $E = 0$ are the leading contribution to the order parameter. These results are fully consistent with the observed enhancement of superconductivity for not too strong disorder and also provide support that the eigenstates that most contribute to the order parameter close to the transition are all critical.

\section{Determination of the critical disorder for the metal-insulator transition by level statistics}\label{sec:levelstatistics}
We have already investigated the interplay of disorder and superconductivity for a broad range of disorder strengths. We have accrued substantial evidence that around $V \sim 12$, the superconducting state undergoes substantial changes. Moreover, the results of the previous section suggest that only a small set of eigenvectors and eigenvalues of the BdG equations contribute substantially to the order parameter. 
Based on these two findings, in this section, we aim to determine the location of the insulating transition with more precision. For this purpose, we carry out an analysis of level statistics of the eigenvalues of the BdG equations. 

We restrict ourselves to the spectral region inside the Debye energy window since our main interest is to characterize the dynamics of the superconducting state. More specifically, we only consider a small set of eigenvalues, from $15$ to $500$ depending on disorder and size, around $E = 0$ which, according to the findings of the previous section, see Fig.~\ref{Fig.puv}, correspond to eigenvectors that contribute substantially to the formation of the order parameter. For those eigenvalues, we compute different spectral correlators: the level spacing distribution and the adjacent gap ratio and its distribution $P(r)$ that characterize quantum dynamics for long times and therefore are sensitive to the insulating transition. We note that in three dimensions, where critical features only occur close to the transition, the superconductor is at the Anderson transition provided that the eigenstates that effectively contribute to the order parameter are all critical. We shall see that this is the case.  

\subsection{The nearest neighbor level spacing distribution $P(s)$}
We note that, in the limit of no disorder, the eigenvalues are two-fold degenerate \cite{Bofan2020}. By turning on disorder, this degeneracy is lifted but for sufficiently weak disorder there is almost no mixing with neighboring eigenvalues. Therefore, the full spectrum is effectively the superposition of two spectra. Since for weak disorder, we expect metallic features, level statistics are expected to be described by the prediction of random matrix theory (Wigner-Dyson statistics). For sufficiently strong disorder, neighboring eigenvalues get mixed and the spectrum is no longer a superposition of two independent spectra. In this case, we still expect agreement with Wigner-Dyson statistics for a single spectrum provided that this system is not too close to the transition. 

\begin{figure}
	\begin{center}
		\subfigure[]{\label{fig.levelstatistic_1} 
			\includegraphics[width=5.6cm]{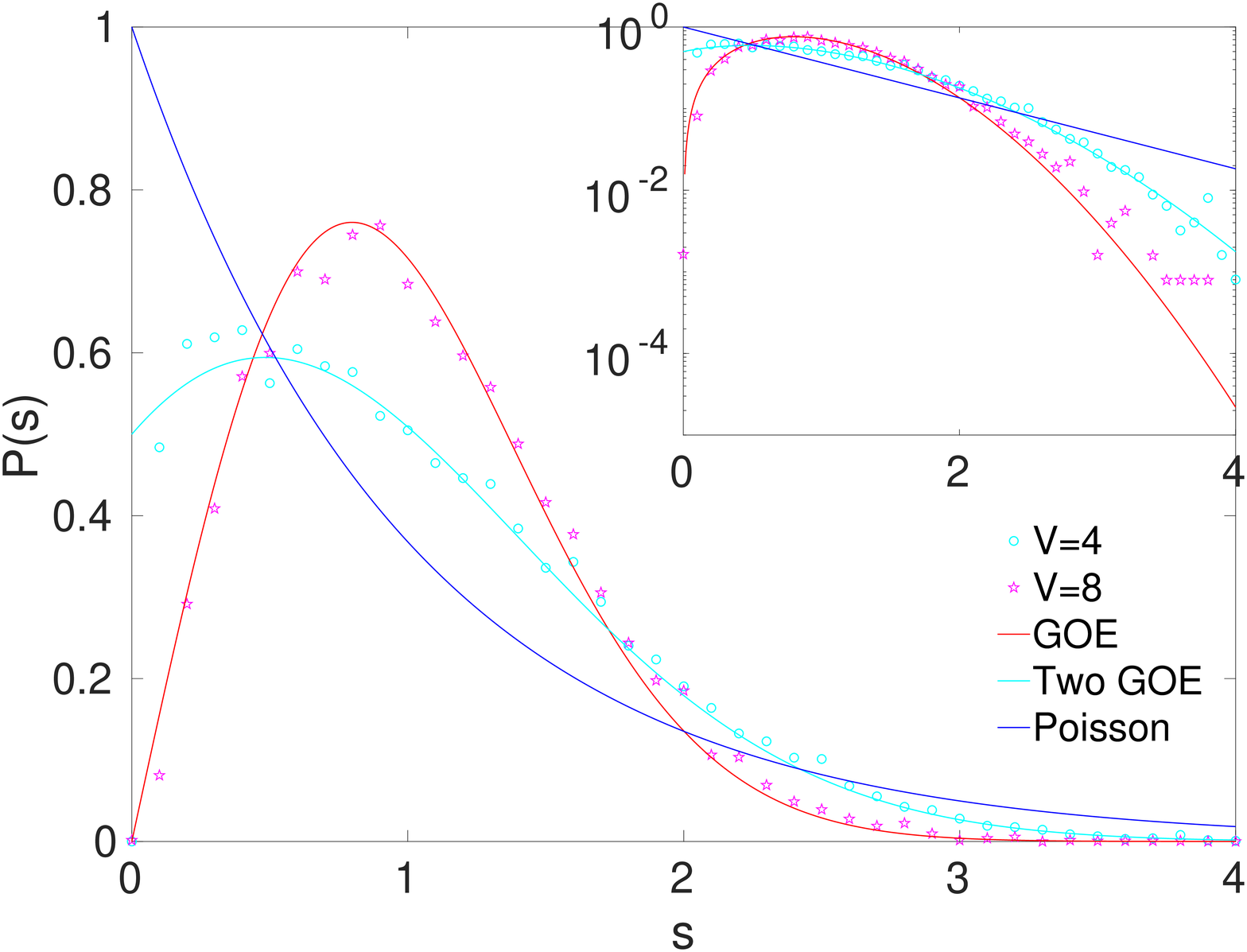}}
		\subfigure[]{\label{fig.levelstatistic_2} 
			\includegraphics[width=5.6cm]{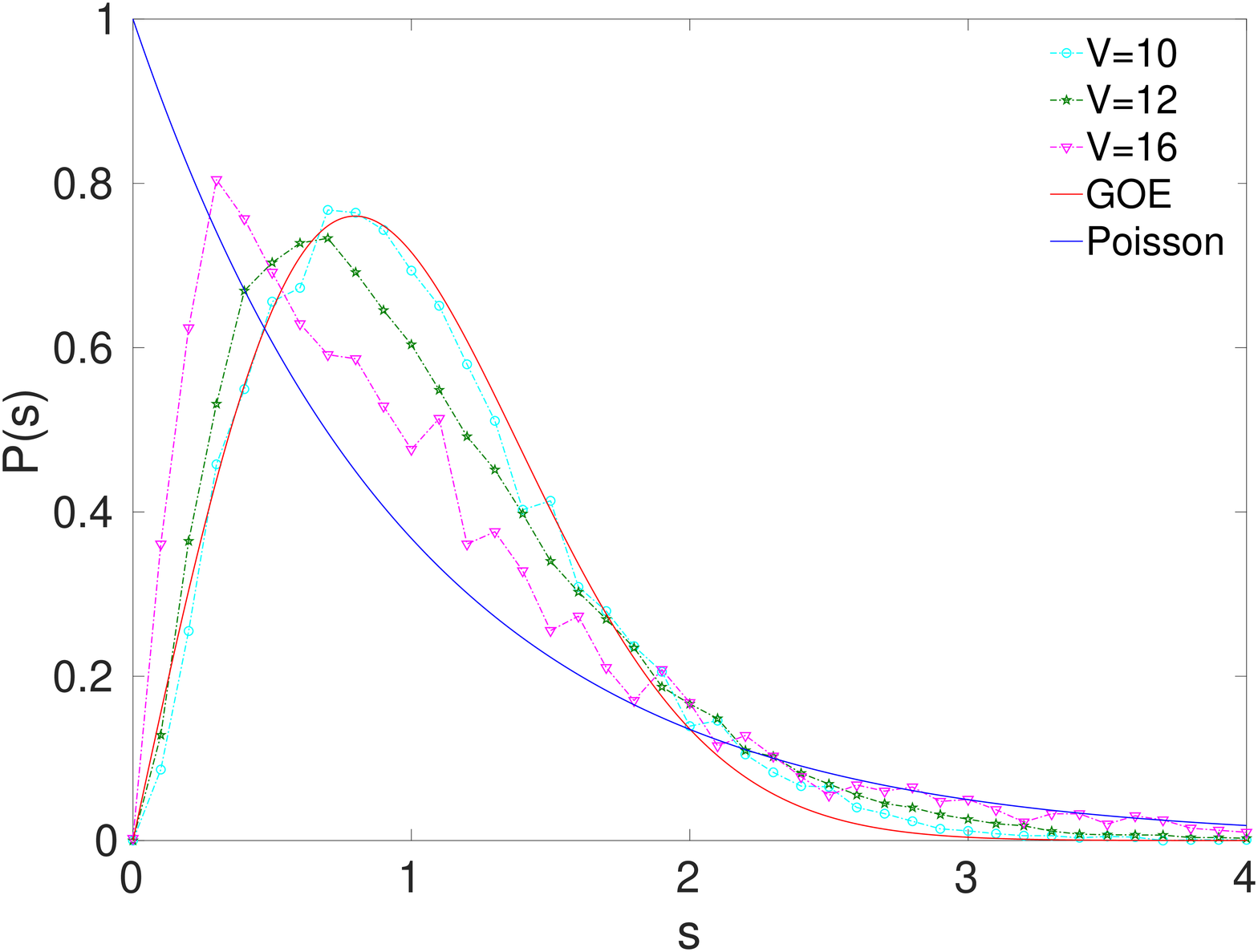}}
		\subfigure[]{\label{fig.levelstatistic_3} 
			\includegraphics[width=5.6cm]{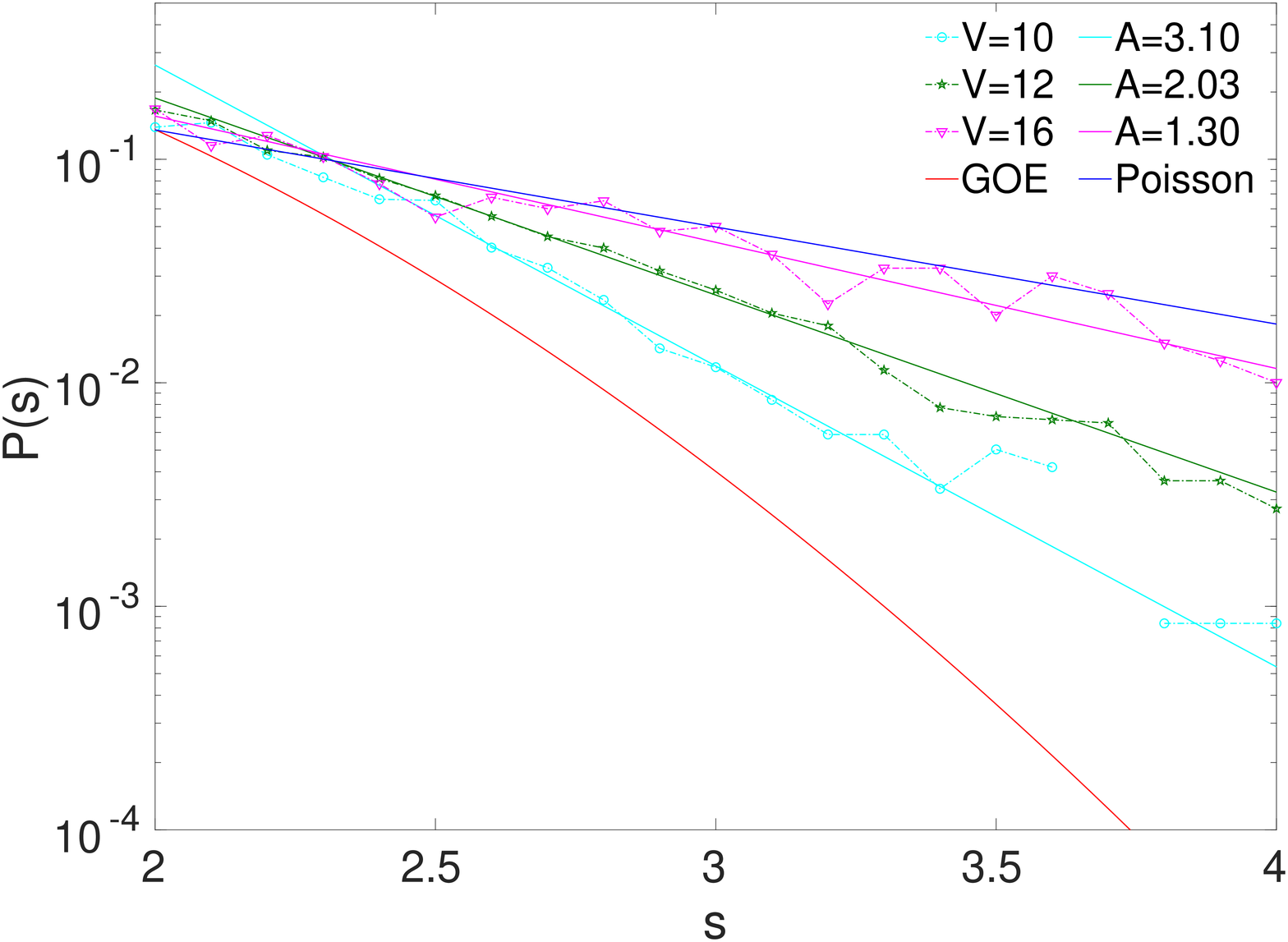}}
		\caption{The nearest neighbor level spacing distribution $P(s)$ for different disorder $V$ for a $20\times 20\times 20$ lattice, a cut-off energy $\omega_D = 2$, $U=-1$ and $\langle n \rangle = 0.875$. \subref{fig.levelstatistic_1}: for weak disorder $V = 4$ we find excellent agreement with the prediction for superposition of two spectra with Wigner-Dyson statistics (two GOEs is indicated by cyan line). Due to the symmetries of the BdG equations, this is the expected result. As disorder increases to $V = 8$, the two spectra are mixed and we observe Wigner-Dyson statistics (GOE is indicated by red line). Inset: Same but in log scale. \subref{fig.levelstatistic_2}: For sufficiently strong disorder $V \geq 10$, the two spectra are mixed and we observe level repulsion typical of a single GOE. For $V = 10$, level statistics are relatively well described by Wigner-Dyson statistics typical of a disordered metal. For $V = 12$, level statistics show typical features of a metal-insulator transition \cite{shapiro1993,varga2000,wang2009} such as level repulsion for $s \ll 1$ and exponential decay for $s \gg 2$. For $V = 16$, $P(s)$ is close to Poisson statistics that characterizes spectral correlations of a disordered insulator. \subref{fig.levelstatistic_3}: The tail of $P(s)$ is fitted by $P(s)= Be^{-As}$, where $A$ and $B$ are the fitting parameters. At $V=12$, the tail of $P(s)$ decays exponentially with $A \approx 2.03$. This is a distinct feature of a system at the Anderson transition.
		}\label{Fig.levelstatistic}
	\end{center}
\end{figure}

Results depicted in Fig.~\ref{Fig.levelstatistic} confirm this picture. For weak disorder, $V = 4$, level statistics agree well with the theoretical prediction for the superposition of two spectra with Wigner-Dyson statistics. The level spacing distribution, namely, the probability of having two consecutive eigenvalues at a distance $s$ in units of the mean level spacing, is in this case \cite{guhr1998} $P_{\mathrm{sup}}(s) = \frac{\pi}{16}s(1-{\rm erf}(\sqrt{\pi}s/4))\exp(-\pi s^2/16)+\frac{1}{2}\exp(-\pi s^2/8)$, where ${\rm erf}(s)$ is the error function. 

As disorder increases $V \sim 8$, we observe that level statistics agree well with the prediction of Wigner-Dyson statistics, also termed the prediction for the Gaussian Orthogonal Ensemble (GOE), but for a single spectrum $P(s)=\frac{\pi}{2}s\exp{(-\frac{\pi s^2} {4})}$, no a superposition \cite{guhr1998}. The reason for that is that a stronger disorder mixes the eigenvalues of the two spectra resulting in a single quantum chaotic spectrum that follows the prediction of random matrix theory  expected in disordered metallic systems. 

For stronger disorder, see Figs,~ \ref{fig.levelstatistic_2} and \ref{fig.levelstatistic_3}, there are deviations from Wigner-Dyson statistics in all spectral correlators: level repulsion is still present in $P(s)$ but the decay is slower than the prediction of Wigner-Dyson statistics. As disorder increases further, it approaches an exponential decay  which is the expectation for a Poisson distribution $P(s) = \exp(-s)$ which characterizes the spectral correlations of disordered insulators.
Around $V = 12$, we observe striking similarities with the spectral features predicted at the Anderson metal-insulator transition \cite{shapiro1993,altshuler1988}. Level repulsion persists but the tail of $P(s)$ decays exponentially $\sim e^{-As}$ with $A > 1$. 
For larger disorder when $V = 16$, the level statistics are close to Poisson statistics which is the expected results for an Anderson insulator. These results suggest a transition around $V = 12$. We will confirm it in next section.

\subsection{The probability distribution of consecutive level spacing $P(r_n)$ and the adjacent gap ratio $\langle \tilde{r}_n \rangle$}
The computation of the $P(s)$ involves the unfolding of the spectrum so that the average mean level spacing is the unity. This process, which in our case was carried out by a low degree polynomials, adds some uncertainty since the results, at least quantitatively, may weakly depend on the unfolding procedure. In order to avoid this problem, we compute the adjacent gap ratio and the distribution of consecutive level spacing that do not require any unfolding. 

The ratio of the consecutive level spacing is defined as \cite{atas2013}
\begin{equation}\label{cls}
r_n = \frac{s_n}{s_{n-1}}
\end{equation}
where $s_n=E_{n+1}-E_{n}$ is the nearest-neighbor spacing of the ordered eigenenergies $E_1 \leq E_2 \leq \dots \leq E_n$. Therefore, the adjacent gap ratio is naturally defined as 
\begin{equation}\label{agr}
\tilde{r}_n = {\mathrm {min}}\left( r_n, \frac{1}{r_n} \right)
\end{equation}
The analytical predictions for the ensemble average of these correlators, and its distributions, for the case of random matrices, that should also apply to quantum disordered metals, is known explicitly \cite{atas2013,Huse2007}. 
A distinct feature of these spectral correlators is its ultra locality, namely, they provide information about time scales much larger than the Heisenberg time. For instance, they provide information about whether the spectrum has (has no) level repulsion as in a metal (insulator). In some sense, it is a zoom in version of the small $s$ limit of $P(s)$. 
For that reason, we expect that  finite size effects, that are more important in this limit, may play some role in suppressing localization effects on the insulating size of the transition. 

We start our analysis with the calculation of the ensemble average adjacent gap ratio $\langle \tilde{r}_n \rangle$ for different disorder $V$. We also carry out a finite size scaling analysis by studying the dependence of the results with $L$. 
 
In order to avoid effects related to the superposition of two spectra, of no interest now, we only consider relatively strong disorder strengths, $V \geq 8$.
  
\begin{figure}
	\begin{center}
		\subfigure[]{\label{fig.adjacent_gap_ratio_1} 
			\includegraphics[width=8.5cm]{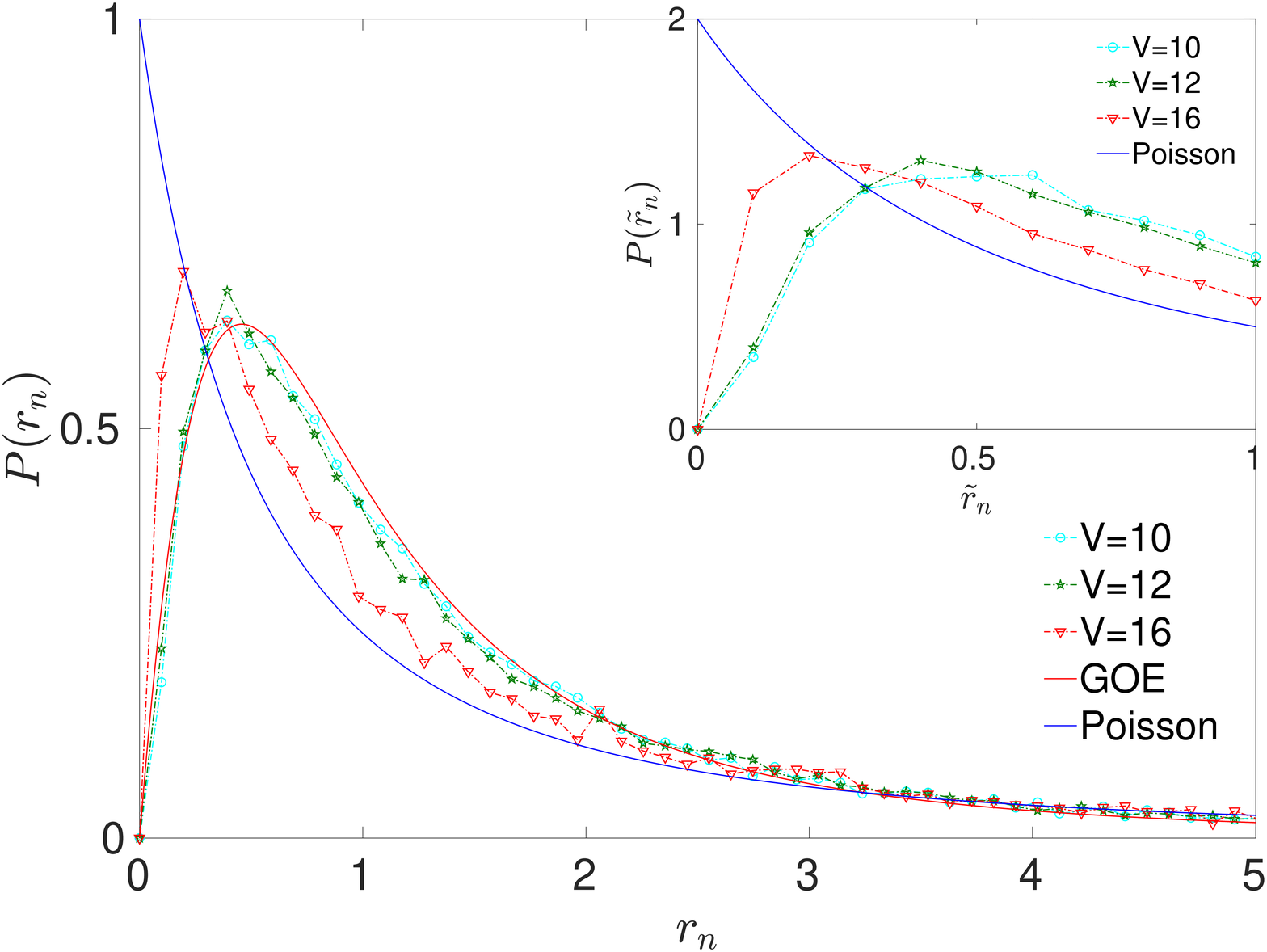}}
		\subfigure[]{\label{fig.adjacent_gap_ratio_2} 
			\includegraphics[width=8.5cm]{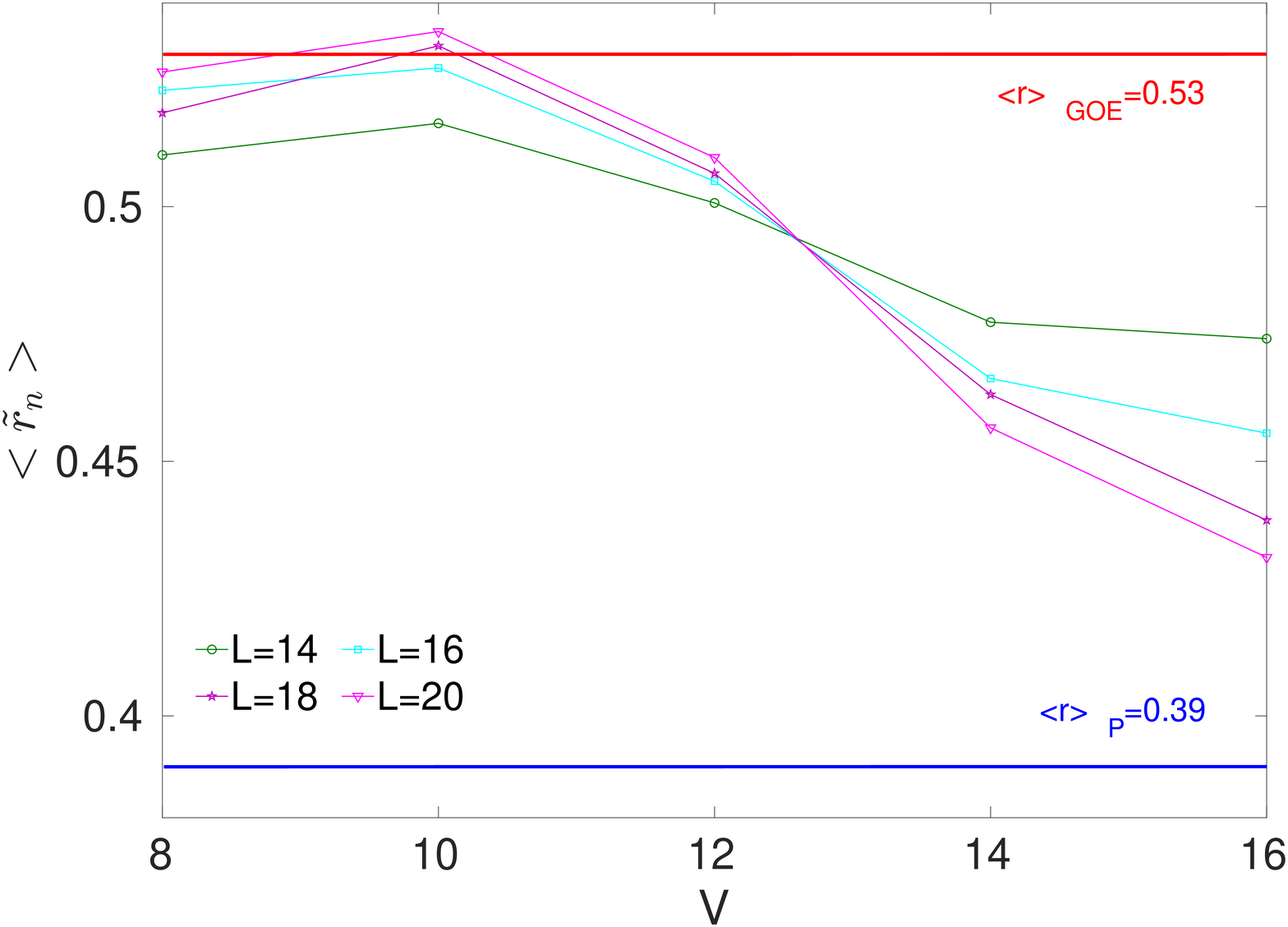}}
		
		\caption{\subref{fig.adjacent_gap_ratio_1}. The probability distribution of the ratio of consecutive level spacing $P(r_n)$, see Eq.~\eqref{cls}. Inset: the probability distribution  of the adjacent gap ratio $P(\tilde{r}_n)$, see Eq.~\eqref{agr}. It shows that even in the transition region $V \sim 12$, the distribution still follows the GOE prediction (red solid line). For $V = 16$, it approaches Poisson statistics. 
		\subref{fig.adjacent_gap_ratio_2}. Finite size scaling analysis of the adjacent gap ratio $\langle \tilde{r}_n \rangle$ as a function of disorder $V$. As disorder increases, we observe a crossover, that becomes sharper as $L$ increases, from the Wigner-Dyson prediction (GOE) that describes the spectral correlations of a disordered metal, to Poisson statistics expected to describe the correlations of a disordered insulator. The crossing point $V_c \approx 12.6$ signals the location of the transition}
		\label{Fig.adjacent_gap_ratio}
	\end{center}
\end{figure}

As is shown in Fig.~\ref{fig.adjacent_gap_ratio_2}, the gap ratio undergoes a crossover from the Wigner-Dyson $\langle \tilde{r}_n \rangle \approx 0.53$ to the Poisson statistics $\langle \tilde{r}_n \rangle \approx 0.39$ \cite{atas2013} around the critical disorder $V \sim 12$. 
More importantly, within the limited range of sizes that we can test numerically, we observe that all curves nicely cross each other at $V \approx 12.6$ so that at this disorder, level correlations are approximately size independent which  is a distinct feature of Anderson transitions \cite{altshuler1988,shapiro1993}. 

For the sake of completeness, we also compute the probability distribution of the ratio of the consecutive level spacing $\langle r_n \rangle$  and the adjacent gap ratio $\langle \tilde{r}_n \rangle$ . We have found, see Fig.~\ref{fig.adjacent_gap_ratio_1}, that even in the critical region $V \approx 12$, the distribution is very close to the Wigner-Dyson  prediction expected in a good disordered metal. Only for much stronger disorder, we observe the transition to Poisson statistics that describes spectral correlations in a disordered insulator. This is not surprising as the adjacent gap ratio is an ultra short-range spectral correlator that is mostly sensitive to level repulsion. The latter is a feature that, because of finite size effects, is still observed in the insulating region not to far from the transition. Indeed, results of the adjacent gap ratio are fully consistent with those of the level spacing distribution. 

In summary, the analysis of spectral correlations, especially the finite size scaling analysis of the adjacent gap ratio, indicates the existence of an Anderson transition around $V_c \approx 12.6$. Level statistics around the transition are intermediate between those of a metal and an insulator and qualitatively similar to those of a three dimensional non-interacting systems at the Anderson transition: level repulsion, a distinctive spectral feature of a disordered metal, is observed but the decay of the level spacing distribution is exponential, as for an insulator $\sim e^{-s}$, though with a larger exponent $\sim e^{-As}$, $A \approx 2$. As disorder increases further, the exponent $A \to 1$ tends to the Poisson statistics result.

\section{Estimation of the critical disorder for the breaking of phase coherence by a percolation analysis}\label{sec:percolation}
We have shown in the previous section that the transition to an insulator occurs around $V_c \sim 12$. A natural question to ask is whether superconducting phase coherence persists until the insulating transition or the loss of global order occurs for weaker disorder. We tentatively address this question by a percolation study of the order parameter. A word of caution is in order, the critical disorder obtained from the percolation analysis is just a rough estimation for the existence, or not, of phase coherence. 

\begin{figure}
	\begin{center}
		\includegraphics[width=10cm]{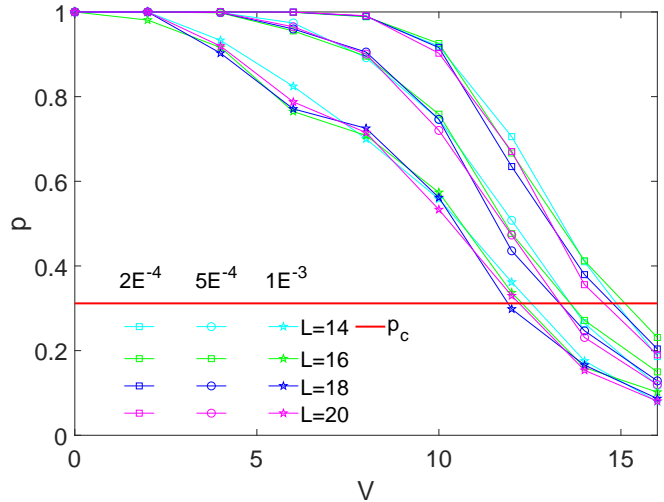}
		\caption{The probability that the amplitude of the order parameter $\langle \Delta(r) \rangle$ is larger than the cut-off value $\Delta_c$ as a function of disorder $V$ and different system sizes $L$. We set three cut-off value $\Delta_c$: $5\times10^{-4}$(circle), $2\times10^{-4}$(square) and $1\times10^{-4}$(triangle). The red line around $0.311$ is the percolation threshold $p_c$ for a simple 3D cubic lattice \cite{stauffer2003introduction}. The interaction term $U=-1$, the Debye energy $\omega_D=2$ and the density $\langle n \rangle = 0.875$.}\label{Fig.percolation}
	\end{center}
\end{figure}

We define that, for a given disorder, the superconductor is phase coherent if the order parameter amplitude $\Delta(r_i)$ forms a percolating cluster. Strictly speaking, a point belongs to the percolating cluster if the order parameter does not vanish. However, on physical grounds, we consider a cut-off value $\Delta_c$ so that if the order parameter is smaller than $\Delta_c$ at a given point, this point does not belong to the percolating cluster. With these assumptions, if the probability $p$ that a point in the sample does not contribute to the percolating cluster is smaller than the percolation threshold $p_c=0.311$ \cite{stauffer2003introduction} for a 3D cubic lattice, then there is no a percolating cluster and phase coherence is lost. Results are shown in Fig.~\ref{Fig.percolation} for different values of the cut-off $\Delta_c$.

As was expected, the location of the transition depends on the chosen cutoff $\Delta_c$. However, the dependence is relatively weak and size independent which allows to estimate with reasonable accuracy, the critical disorder $ V_c \approx 13 \pm 1$ at which the percolation transition occurs. Interestingly, it is very close at the critical disorder at which the insulating transition takes place. Although further research would be necessary, such as an explicit calculation of the superfluid density, to settle this issue, our findings suggests that phase coherence may be lost around the same range of disorder at which the insulating transition occurs. 

In summary, both the percolation and the insulating transition take place at a similar disorder strength. 
Although the percolation analysis does not provide a precise determination of the critical disorder for the loss of phase coherence, this fact suggests that phase coherence is likely lost at a similar value of disorder.

\section{Discussion and conclusions}\label{sec:discussion}
The results of the paper, together with previous findings in two dimensions, provide a rather detailed picture of the interplay of disorder and superconductivity, especially in the critical region around the Anderson transition:

First, it is beyond any reasonable doubt that disorder does affect profoundly the superconducting state. The amplitude of the order parameter, even in the metallic region and relatively far from the transition, has a broad spatial distribution. Close to the transition is log-normal, at least in the range of sizes we can test, in both 2D and 3D. 
The singularity-spectrum, related to the amplitude distribution of the order parameter, is parabolic as that of the density of multifractal eigenstates at the Anderson transition. This emerging picture, seem to disagree with the predictions of the Anderson theorem that disorder does not affect qualitatively the superconducting state. However, we consider it disagrees with the many interpretations of the Anderson theorem in the literature rather than with the original content of Anderson's statement \cite{Anderson1959}.

Second, the answer to the question about whether disorder can enhance superconductivity is responded affirmatively. In both 2D and 3D, this enhancement occurs for a broad range of disorder strengths but only for weak electron-phonon coupling. The averaged order parameter could be enhanced up to two or three times, especially in 2D. However, it is likely that the enhancement of the critical temperature will be much less due to phase fluctuations induced by disorder. Therefore, it is uncertain that disorder can enhance the global critical temperature to the point that it is relevant for practical applications. Likewise, in 3D, the maximum enhancement occurs around the transition, a region where thermal and quantum fluctuations, that lower the critical temperature, will be larger. Therefore, it is unclear to what extent this enhancement of the order parameter is also observed in the critical temperature. This perception could change with the discovery of a weakly coupled superconducting material with a critical temperature above the one for MgB$_2$. 

Third, despite of the strong spatial fluctuations, phase coherence holds approximately until the critical disorder at which the insulating transition occurs. 

Fourth, all quantum coherence effects, from the strength of spatial fluctuations to the enhancement of superconductivity of the order parameter, become more prominent as either the electron-phonon coupling strength or the Debye energy decreases.  

Fifth, natural extensions of this research include the effect of Coulomb interaction and a perpendicular magnetic field. Regarding the former, charging effects could be included by assuming that the inhomogeneities could be seen as a Josephson junction array where the introduction of charging effects is simpler. Regarding the latter, it would be interesting to investigate different aspects of vortexes physics and, in special, the Kosterlitz-Thouless transition in a superconducting state with multifractal-like features. Likewise, the study of finite temperature effects and transport properties around the Anderson transition are others natural extensions of this work. We aim to address some of these problems in the near future. 

In conclusion, we have investigated the superconducting state around the Anderson transition that in the non-interacting limit is described by multifractal eigenstates by using the BdG formalism. We have found that the spatial average of the order parameter is enhanced as disorder is increased but only for disorder strength below the transition. The distribution of the order parameter is log-normal around the transition. For lower disorder, it is still broad and asymmetric that illustrate the important role of disorder even relatively far from the transition. As for non-interacting electrons at the Anderson transition, the singular spectrum is parabolic and level statistics are intermediate between Poisson and random matrix theory predictions. All these are typical features of systems where multifractality plays an important role. A qualitative percolation analysis reveals that the loss of phase coherence is likely to occur at around the same disorder as the superconductor-insulator transition.  
\acknowledgments  We acknowledge financial support from a Shanghai talent program and from the National Natural Science Foundation of China (NSFC) (Grant number 11874259)

\bibliographystyle{unsrt}
\bibliography{library} 
\end{document}